\documentclass[12pt]{article}

\usepackage{graphicx}
\usepackage{alltt}
\usepackage{amsmath}
\usepackage{amssymb}

\usepackage{amsmath}
\usepackage{hyperref}
\usepackage{graphics}

\usepackage{authblk}

\newcommand{\capzero}{\stackrel{0}{\cap}}
\newcommand{\cupzero}{\stackrel{0}{\cup}}

\def\upar{\mbox{\tiny  \bf up}}
\def\dnar{\mbox{\tiny  \bf dn}}

\newcommand {\SO}{\mathrm{SO}}
\newcommand {\sso}{\mathfrak{so}}
\newcommand {\C}{\mathbb{C}}
\newcommand{\beq}{\begin{equation}}
\newcommand{\eeq}{\end{equation}}
\newcommand{\beqr}{\begin{eqnarray}}
\newcommand{\eeqr}{\end{eqnarray}}

\title{BRST cohomology of the sum of two pure spinors}
\author[*]{Andrei Mikhailov}
\author[**]{Renjun Xu}
\affil[*]{Instituto de F\'{i}sica Te\'orica, Universidade Estadual Paulista\\
 R. Dr. Bento Teobaldo Ferraz 271,
 Bloco II -- Barra Funda\\
 CEP:01140-070 -- S\~{a}o Paulo, Brasil}
\affil[**]{Department of Physics, University of California \\ Davis, CA 95616, USA}

\begin{document}

\begin{titlepage}

% \begin{center}
% {\Large\bf $ $ \\ $ $ \\
% BRST cohomology of the sum of two pure spinors
% }\\
% \bigskip\bigskip\bigskip
% {\large Andrei Mikhailov} and {\large Renjun Xu}
% \\
% \bigskip\bigskip
% {\it Instituto de F\'{i}sica Te\'orica, Universidade Estadual Paulista\\
% R. Dr. Bento Teobaldo Ferraz 271,
% Bloco II -- Barra Funda\\
% CEP:01140-070 -- S\~{a}o Paulo, Brasil
% }\\
% \bigskip\bigskip\bigskip

% \\
% \bigskip\bigskip
% {\it Renjun please fill your data here}

% \vskip 1cm
% \end{center}

\maketitle

\begin{abstract}
We study the zero mode cohomology of the sum of two pure spinors. The knowledge
of this cohomology allows us to better understand the structure of the massless
vertex operator of the Type IIB pure spinor superstring.
\end{abstract}

\end{titlepage}

\tableofcontents

\section{Introduction}
Pure spinor formalism is perhaps the most promising approach to the high loop
calculations in superstring theory. But in fact, it is even useful for the
classical supergravity \cite{Berkovits:2001ue}. The pure spinor description of the Type IIB SUGRA
is somewhat analogous to the pure spinor description of the supersymmetric
Yang-Mills theory. But it is very different in details, and in fact much less
understood.

In particular, we need to better understand the {\em massless} vertex
operators. They are the cohomology classes describing the infinitesimal
deformations of the given SUGRA solution.

In flat space-time, the massless vertex operators are {\em roughly speaking}
products of the expression built on the left-movers, and the expression built
from the right-movers. These left- and right-handed parts are very similar to
the Maxwell vertices. This, however, does not work in curved space-time, where
the separation into left- and right-movers does not exist. And even in flat
space-time, there are subtleties at the near-zero momentum \cite{Mikhailov:2012id,Mikhailov:2012uh}.

A construction of the massless vertex in $AdS_5\times S^5$ was suggested in \cite{Mikhailov:2011af}.
Unfortunately it involved rather complicated calculations. In this paper we
will get a better understanding of the construction, by setting up a
slightly different cohomological perturbation scheme. The main component of
this new scheme is the following ``zero mode'' BRST operator:
\begin{equation}
   Q^{(0)} =  \left(\lambda_L^{\alpha} + \lambda_R^{\alpha}\right)
{\partial\over\partial \theta^{\alpha}}
\end{equation}
where $\lambda_L^{\alpha}$ and $\lambda_R^{\alpha}$ are both pure spinors, {\it i.e.} satisfy the constraints:
\begin{equation}
   (\lambda_L\Gamma^m\lambda_L) = (\lambda_R\Gamma^m\lambda_R) = 0
\end{equation}
and $\theta^{\alpha}$ are free Grassmann variables. It is assumed that $Q^{(0)}$ acts on
{\em polynomials} of $\lambda_L^{\alpha},\lambda_R^{\alpha},\theta^{\alpha}$. This cohomology problem was first suggested
in \cite{BerkovitsIPMU}.

This $Q^{(0)}$ is the zeroth approximation to the BRST operator in $AdS_5\times S^5$,
acting on the chiral state. It turns out that the structure of $H(Q^{(0)})$ is
rather rich. The full BRST operator induces on $H(Q^{(0)})$ the nilpotent
operator $d_1$, then on the cohomologies of $d_1$ we get $d_2$, {\it etc.}  The
resulting spectral sequence converges to the Type IIB vertex operators. To
understand the construction of vertex operators, the following facts about the
cohomology of $Q^{(0)}$ are useful:
\begin{enumerate}
\item There is a class $\simeq \lambda^2\theta^2$; the vertex is built by multiplying this
   class by some function $f(x)$
\item There is a class of the order $\lambda^3\theta^3$, but its quantum numbers do not match
   the quantum numbers of the potential obstacle; therefore $d_1$
   annihilates the vertex
\item However, there is a nontrivial class of the order $\lambda^3\theta^5$.
   This implies that $d_2$ of the vertex is potentially nonzero. In fact
   vanishing of $d_2$ requires that $f(x)$ is a harmonic function. This is the
   expected on-shell condition.
\end{enumerate}
Another construction of the massless vertex, which emphasizes the 
boundary-to-bulk structure, was suggested in \cite{BerkovitsIPMU,Berkovits:2012ps}. For that construction, the
structure of the cohomology of $Q^{(0)}$ is also important.

\section{Vertices from parabolic induction}
In this section we will review the construction of pure spinor vertex
operators in $AdS_5\times S^5$, and explain how the knowledge of the cohomology
of $Q^{(0)}$ allows to better organize the calculations.

\subsection{Lie superalgebra $psl(4|4)$}
\paragraph     {Structure of the superalgebra}
The even subalgebra of $sl(4|4)$ is a direct sum of two Lie algebras\footnote{The real form used in AdS/CFT is ${\bf g}_{\upar} = {\bf su(2,2)}$ and ${\bf g}_{\dnar} = {\bf su(4)}$.}:
\begin{equation}
{\bf g}_{even} = {\bf g}_{\upar} \oplus {\bf g}_{\dnar}
\end{equation}
Schematically, in the $4\times 4$-block notations:
\begin{equation}
{\bf g} = \left[ \begin{array}{cc}
      {\bf g}_{\upar}  & {\bf n}_+ \cr
      {\bf n}_-         & {\bf g}_{\dnar}
\end{array}\right]
\end{equation}
Notice that $\bf n_+$ and ${\bf n}_-$ are both odd abelian subalgebras ${\bf C}^{0|16}$.

\paragraph     {Super coset space $AdS_5\times S^5$}
There is a ``denominator'' subalgebra:
\begin{equation}
{\bf g}_{\bar{0}}= sp(2)\oplus sp(2)\;\subset\; {\bf g}_{\rm even}
\end{equation}
The AdS space is the coset space \cite{Roiban:2000yy}:
\begin{equation}
   AdS_5\times S^5 = {PSL(4|4)\over Sp(2)\times Sp(2)}
\end{equation}
The subalgebra ${\bf g}_{\bar{0}}$ is the stabilizer of a point in $AdS_5\times S^5$. We will denote
$u^{\alpha}$ the elements of the fundamental representation of ${\bf g}_{\rm up} = sl(4)$ and $v_a$ the
elements of the anti-fundamental of ${\bf g}_{\rm dn} = sl(4)$. Being a direct sum of two
symplectic algebras, ${\bf g}_{\bar{0}}$ can be characterized as a stabilizer of a pair of
symplectic forms  $\omega^{\alpha\beta}$ and $\omega_{ab}$ \cite{Roiban:2000yy}.
(With a slight abuse of notations, we use the same letter $\omega$ for both of them.)
We introduce some additional notations, see also Section \ref{sec:NotationsAndAbbreviations}:
\begin{equation}\label{FunnyNotations}
   u\cup v = u^{\alpha}\omega_{\alpha\beta}v^{\beta}\;,\;\;
   u\cap v = u_a \omega^{ab} v_b\;,\;\;
   ||u|| = u^{\alpha\beta}\omega_{\beta\alpha} \mbox{ \tt\small or }
   u_{ab}\omega^{ba}
\end{equation}

\subsection{Parabolic induction}
The idea of \cite{Mikhailov:2011af} is to first construct the vertex for the ``chiral'' states
--- those states  which are annihilated by all the elements of  ${\bf n}_-$. Then,
having the vertex for the chiral states, we can obtain the general vertex by
applying the $PSL(4|4)$ rotations. Let us introduce the following coordinates
on the group manifold $PSL(4|4)$:
\begin{equation}\label{ChoiceOfCoordinates}
   g = e^{\omega}e^{\theta_+} e^x e^{\theta_-}
\end{equation}
where $\omega\in {\bf g}_{\bar{0}}$, $\theta_{\pm}\in {\bf n}_{\pm}$, and $x$ is in the complement of $\omega$ in ${\bf g}_{\rm even} = {\bf g}_{\bar{2}} + {\bf g}_{\bar{0}}$.

For the chiral state, the vertex:
\begin{itemize}
   \item will not depend on $\theta_-$
   \item will transform in a given representation $L$ of ${\bf g}_{\rm even}$ under the shifts
      of $x$
\end{itemize}
We therefore use the {\em parabolic induction} from the following parabolic
subalgebra:
\begin{equation}\label{ParabolicSubalgebra}
{\bf p} = \left[ \begin{array}{cc}
      {\bf g}_{\upar}    &  0 \cr
      {\bf n}_-         & {\bf g}_{\dnar}
\end{array}\right]
\end{equation}
We start with a representation $L$ of the {\em bosonic} Lie algebra
${\bf g}_{\rm even} = {\bf g}_{\upar} \oplus {\bf g}_{\dnar}$. We can only apply our construction in the case when $L$
satisfies the following properties:
\begin{enumerate}
   \item The quadratic Casimir of ${\bf g}_{\rm even}$ vanishes on $L$
   \item In a dual representation\footnote{For any representation $L$, the
{\em dual} representation $L'$ is on the space of all linear functions on $L$} $L'$, exists a vector $\Omega\in L'$ such that the
      subalgebra ${\bf g}_{\bar{0}}\subset {\bf g}_{\rm even}$ annihilates $\Omega$.
\end{enumerate}

\paragraph     {Kac module. Chiral and non-chiral elements.}
We will construct some vertex for every element of the Kac module:
\begin{equation}
   \mbox{Ind}_{\bf p}^{\bf g} L = {\rm U}{\bf g}\otimes_{\bf p} L
\end{equation}
\paragraph     {Definition of chiral {\it vs.} non-chiral}
Those elements of $\mbox{Ind}_{\bf p}^{\bf g} L$ which are of the type ${\bf 1}\otimes l$, where ${\bf 1}$ is the unit
of ${\rm U}{\bf g}$ and $l$ is a vector in $L$, will be called {\em chiral} elements. All
other elements will be called {\em non-chiral}.

It was argued in \cite{Mikhailov:2011af} that most of elements of $\mbox{Ind}_{\bf p}^{\bf g} L$ give by our construction
BRST-trivial vertices\footnote{this follows from the consistency of AdS/CFT,
and also from identifying the R-charge of the Type IIB SUGRA; chiral states
have R-charge $+2$, lowering operators decrease the R-charge, the R-charge
cannot be less than $-2$}. But all chiral and some non-chiral elements give
BRST-nontrivial vertices.

\paragraph     {Definition of $v(\theta_+,\lambda)$}
In order to construct the vertex, we first define  $v(\theta_+,\lambda)\in L'$ --- a
function of $\theta_+$ and $\lambda$ taking values in $L'$. It is defined
in terms of $\Omega$:
\begin{equation}\label{VvsOmega}
v(\theta_+,\lambda) = \;
||\lambda_{R+}\cup\theta_+\cap\lambda_{L+}\cup\theta_+||\;
 \Omega + \ldots
\end{equation}
where $\ldots$ stand for the terms of the higher order in $\theta_+$ which we will define
later.

\paragraph     {Vertex for chiral states.}
The ansatz for the massless vertex corresponding to the {\em chiral} state
${\bf 1}\otimes l$ is, using the coordinates defined in (\ref{ChoiceOfCoordinates}):
\begin{align}\label{VertexForChiral}
   V(\theta_+,x,\lambda) \;=\;& \langle v(\theta_+,\lambda), e^xl \rangle
\end{align}

\paragraph     {Vertex for non-chiral states.}
For a general ({\it i.e.} non-chiral) state:
\[(\eta_{1+}\cdots\eta_{k+})\otimes l\]
the vertex $V(\theta_+,\theta_-,x,\lambda)$ can be calculated in the following way. Remembering
that $g = e^{\omega}e^{\theta_+} e^x e^{\theta_-}$, we write:
\begin{align}
   V(\theta_+,\theta_-,x,\lambda) = \;&
   \langle v(\theta_+,\lambda), \;\;g\;\;\;
   (\eta_{1+}\cdots\eta_{k+})\otimes l \rangle \; =
\nonumber \\ 
=\; &
   \langle v(\theta_+,\lambda), \;e^{\omega}e^{\theta_+} e^x e^{\theta_-}\;
   (\eta_{1+}\cdots\eta_{k+})\otimes l \rangle \;
\end{align}
then expand in $\theta_-$ and pass all the $\theta_-$ through $\eta_+$'s until they all are eaten
into the commutators with $\eta_+$ resulting in elements of ${\bf g}_{\rm even}$, which then
rotate $l$ as $l\in L$ --- a representation of ${\bf g}_{\rm even}$. In other words:
\begin{itemize}
\item vertex operators for non-chiral states are obtained from the vertex
operators for the chiral state (\ref{VertexForChiral}) by applying supersymmetry transformations
\end{itemize}
The action of the BRST operator in these notations is, schematically:
\begin{align}
   Q = \;& Q_L + Q_R
\\ 
\epsilon Q_L v = \; &
\left(
   \epsilon \lambda_{L+}  +
   (\theta_+\cap\epsilon\lambda_{L+}\cup\theta_+)
\right){\partial\over\partial\theta_{+}}\;v
\;+ \; (\epsilon\lambda_{L+} \Gamma^m \theta_+)t_m^2v
\label{FullQL}
\\ 
\epsilon Q_R v = \; &
\left(
   \epsilon \lambda_{R+}  -
   (\theta_+\cap\epsilon\lambda_{R+}\cup\theta_+)
\right){\partial\over\partial\theta_{+}}\;v
\;- \; (\epsilon\lambda_{R+} \Gamma^m \theta_+)t_m^2v
\label{FullQR}
\end{align}
where $t_m^2$ are generators of ${\bf g}_{\bar{2}}$. Details are in \cite{Mikhailov:2011af}.

\subsection{Perturbation theory}
Let us develop a perturbation theory considering $\theta_+,\;\lambda_{L+},\;\lambda_{R+}$ as small of
the same order $\varepsilon$. In other words, let us consider the expansion of $v$ in
powers of $\theta_+$. The zeroth order approximation to $Q$ is:
\begin{equation}\label{QSum}
   Q^{(0)} = (\lambda_{L+} + \lambda_{R+}){\partial\over\partial\theta_+}
\end{equation}
Therefore to understand the structure of the vertex we need to start with
calculating the cohomology of (\ref{QSum}). We observe the following facts.
 
\paragraph     {Order $\lambda^2\theta^2$}
The cohomology of $Q^{(0)}$ at the order $\lambda^2\theta^2$ is generated by the coefficient of
$\Omega$ in (\ref{VvsOmega}):
\begin{equation}
 \Phi_{\rm scalar}^{[\lambda^2\theta^2]} =  ||\lambda_{R+}\cup\theta_+\cap\lambda_{L+}\cup\theta_+||
\end{equation}
and some $\Phi^{[\lambda^2\theta^2]}_{\small\rm 2-form}$.

The second approximation $Q^{(1)}$ brings $||\lambda_{R+}\cup\theta_+\cap\lambda_{L+}\cup\theta_+||\;
 \Omega$ to something
of the order $\lambda^3\theta^3$.

\paragraph     {Order $\lambda^3\theta^3$}
The cohomology of $Q^{(0)}$ at the order $\lambda^3\theta^3$ is nonzero. However, it does not
create an obstacle to completing the terms of the order $\lambda^2\theta^4$ in $v(\theta_+,\lambda)$. This
is because, even being nonzero, its quantum numbers ({\it i.e.} representation
content under ${\bf g}_{\bar{0}}$) do not match the quantum numbers of a potential obstacle.
Indeed, the question is whether we could cancel the last term in (\ref{FullQL})
and (\ref{FullQR}), {\it i.e.} the terms proportional to $t_m^2\Omega$. But $\Omega$ is ${\bf g}_{\bar{0}}$-invariant,
therefore $t_m^2\Omega$  transforms as a vector of the tangent space to sphere, plus a
vector of the tangent space to AdS. But looking at either (\ref{PhiL3T3Up}), (\ref{PhiL3T3Dn}), (\ref{PhiL3T3Mixed1})
and (\ref{PhiL3T3Mixed2}), or at (\ref{PhiL3T3}), we see that there is no class of the type $\lambda^3\theta^3$ with
such quantum numbers.
 
Therefore
there is no obstacle to completing the terms of the order $\lambda^2\theta^4$ in $v(\theta_+,\lambda)$.

\paragraph     {Order $\lambda^3\theta^5$}
But the action of $Q^{(1)}$ on these terms of the order $\lambda^2\theta^4$ produces terms of the
order $\lambda^3\theta^5$. It turns out that there is precisely one cohomology class at the
order $\lambda^3\theta^5$. But the contribution of this term to $Qv$ is proportional to the
action of the Casimir of ${\bf g}_{\rm even}$ on $\Omega$, which is zero by the assumption.

\paragraph     {Order $\lambda\theta$}
There are two nontrivial cohomology classes at the order $\lambda\theta$:
\begin{align}
\Phi^{[\lambda\theta]}_{\rm up} = \;&
\lambda_{L+}\capzero \theta_+ + \theta_+\capzero \lambda_{L+} -
(\theta_+\capzero \lambda_{R+} + \lambda_{R+}\capzero\theta_+)
\\ 
\Phi^{[\lambda\theta]}_{\rm dn} = \;&
\lambda_{L+}\cupzero \theta_+ + \theta_+\cupzero \lambda_{L+} -
(\theta_+\cupzero \lambda_{R+} + \lambda_{R+}\cupzero\theta_+)
\end{align}
Is it possible that $v = Qw$ where $w$ is composed of $\Phi^{[\lambda\theta]}_{\rm up}$ and $\Phi^{[\lambda\theta]}_{\rm dn}$ plus
terms of the higher order in $\theta$? As explained in \cite{Mikhailov:2011af}, this is not possible for
the following reason. Consider the transformation $E$ which exchanges
$\lambda_{L+} \leftrightarrow \lambda_{R+}$ and multiplies $\theta_+$ by $i$:
\begin{equation}\label{ESymmetry}
   E\lambda_{L+} = \lambda_{R+}\;,\;
   E\lambda_{R+} = \lambda_{L+}\;,\;
   E\theta_+ = i\theta_+
\end{equation}
This is a symmetry of the BRST complex, in the following sense:
\begin{equation}
   EQ = -i QE
\end{equation}
We observe that $E\Phi_{\rm up}^{[\lambda\theta]} = -i \Phi_{\rm up}^{[\lambda\theta]}$ and $E\Phi_{\rm dn}^{[\lambda\theta]} = -i \Phi_{\rm dn}^{[\lambda\theta]}$, therefore
$EQ\Phi^{[\lambda\theta]} = - Q\Phi^{[\lambda\theta]}$. At the same time:
\begin{equation}\label{ETLTM}
E\;||\lambda_{R+}\cup\theta_+\cap\lambda_{L+}\cup\theta_+|| =
||\lambda_{R+}\cup\theta_+\cap\lambda_{L+}\cup\theta_+||
\end{equation}
This means that $v$ cannot be cancelled by $\Phi^{[\lambda\theta]}_{\rm up}$ and $\Phi^{[\lambda\theta]}_{\rm dn}$.

Is it possible that $v = Qw$ where $w$ is not chiral
({\it i.e.} depends on both $\theta_+$ and $\theta_-$)? This depends on the properties
of $L$. For the type of problems usually encountered in AdS/CFT correspondence
$L$ is a tensor product of some representation of ${\bf g}_{\rm up} = so(2,4)$ and a
{\em finite-dimensional} representation of ${\bf g}_{\rm dn} = so(6)$. Suppose that the
quantum numbers of this finite-dimensional representation of $so(6)$ are large
enough. Then \cite{Mikhailov:2009rx} implies that, when $v = Qw$ and $v$ transforms in $\mbox{Coind}_{\bf p}^{\bf g} L'$,
then $w$ can also be chosen to transform in the same representation
$\mbox{Coind}_{\bf p}^{\bf g} L'$. Therefore, if $v$ represents a nontrivial cohomology class in the
chiral BRST complex, then it also represents a nontrivial cohomology class
in the full BRST complex.

\section{Cohomology of the sum of two pure spinors.}\label{sec:CohomologyOfTheSum}
In this section we will give the result for  the cohomology of the nilpotent
operator:
\begin{equation}\label{DefQ0}
   Q^{(0)} = (\lambda_{L+} + \lambda_{R+}){\partial\over\partial\theta_+}
\end{equation}
where $\lambda_{L+}$ and $\lambda_{R+}$ are two different pure spinors, acting on the polynomials
of $\lambda_L,\lambda_R,\theta$, up to the ghost number 3. The derivation of this results will
follow in the next sections.

\subsection{Table of cohomology}\label{sec:FlatSpaceNotations}
Up to the ghost number three, we find the following cohomology classes.
Smallcase latin letters enumerate the basis vecotrs in ${\bf C}^{10}$, and capital
letters are spinor indices.

\paragraph     {Ghost number 1}
\begin{align}
\Phi^{[\lambda]} = \;& \lambda_{L} - \lambda_{R}
\\ 
\left(\Phi^{[\lambda\theta]}\right)^p = \;&
\left((\lambda_L - \lambda_R)\Gamma^p \theta\right)
\end{align}

\paragraph     {Ghost number 2}
\begin{align}
\left(\Phi^{[\lambda^2]}\right)^{pqrst} = \;&
\left((\lambda_L - \lambda_R)\Gamma^{pqrst}_{AB}
   (\lambda_L - \lambda_R)\right)
\\
\left(\Phi^{[\lambda^2\theta]}\right)^{pA} = \;&
((\lambda_L - \lambda_R)\Gamma^p\theta)\; (\lambda^A_L-\lambda^A_R)
\mbox{ \tt\small mod }\ldots
\\ 
\Phi_{\tiny\tt scalar}^{[\lambda^2\theta^2]} = \;&
(\lambda_L\Gamma^p\theta)(\lambda_R\Gamma^p\theta)
\\ 
\left(\Phi_{\tiny\tt 2-form}^{[\lambda^2\theta^2]}\right)_{pq} = \;&
(\theta\Gamma_{r_1r_2r_3}\theta)(\lambda_L\Gamma_{pqr_1r_2r_3}\lambda_R)
- 18(\theta\Gamma_{pqr}\theta)(\lambda_L\Gamma^r\lambda_R)
\label{PhiLLTT2Form}
\end{align}

Notice that all classes of the ghost number 2 except $\Phi_{\tiny\tt scalar}^{[\lambda^2\theta^2]}$ are even under
$\lambda_L\leftrightarrow\lambda_R$; the class $\Phi_{\tiny\tt scalar}^{[\lambda^2\theta^2]}$ is odd under $\lambda_L\leftrightarrow\lambda_R$.

\paragraph     {Ghost number 3}
\begin{align}
\left(\Phi^{[\lambda^3]}\right)^{ABC} = \;&
(\lambda^A_L-\lambda^A_R)\;(\lambda^B_L-\lambda^B_R)\;(\lambda^C_L-\lambda^C_R)
\mbox{ \tt\small mod }\ldots
\\[5pt]
\left(\Phi^{[\lambda^3\theta]}\right)^{pAB} = \;&
((\lambda_L - \lambda_R)\Gamma^p\theta)\; (\lambda^A_L-\lambda^A_R)
\;(\lambda^B_L-\lambda^B_R)
\mbox{ \tt\small mod }\ldots
\\[5pt]
\left(\Phi^{[\lambda^3\theta^2]}\right)^A_{pq} = \;&
(\lambda^A_L- \lambda_R^A)\;
\left(\Phi_{\tiny\tt 2-form}^{[\lambda^2\theta^2]}\right)_{[pq]}
\nonumber \\
\;&\mbox{ \tt\small mod }\;
(\Gamma_{pq}\Gamma^{rs}(\lambda_L- \lambda_R))_A\;
\left(\Phi_{\tiny\tt 2-form}^{[\lambda^2\theta^2]}\right)_{[rs]}
\label{Lambda3Theta2}\\[5pt]
\Phi_{pqr}^{[\lambda^3\theta^3]} = \;&
\Gamma_{pqr}^{AB}{\partial\over\partial\theta^A}{\partial\over\partial\theta^B}
\Phi^{[\lambda^3\theta^5]}
\label{PhiL3T3}\\[5pt]
\Phi^{[\lambda^3\theta^4]}_A = \;& {\partial\over\partial\theta^A}
\Phi^{[\lambda^3\theta^5]}
\label{TypeL3T4Gamma}
\\[5pt]
\Phi^{[\lambda^3\theta^5]} = \;&
\left(1 - {5\over 3}\left(\lambda_L{\partial\over\partial\lambda_R}\right)\right)
(\lambda_R\Gamma^p\theta)(\lambda_R\Gamma^q\theta)(\lambda_R\Gamma^r\theta)
(\theta\Gamma_{pqr}\theta) \;-
\nonumber \\ 
\;&
- \left(1 - {5\over 3}\left(\lambda_R{\partial\over\partial\lambda_L}\right)\right)
(\lambda_L\Gamma^p\theta)(\lambda_L\Gamma^q\theta)(\lambda_L\Gamma^r\theta)
(\theta\Gamma_{pqr}\theta)
\label{TypeL3T5Gamma}
\end{align}
In Eq. (\ref{TypeL3T5Gamma}) the notation $\lambda_R{\partial\over\partial\lambda_L}$ stands for a formal differentiation w.r.to
$\lambda$, without taking into account the pure spinor constraint. In other words, it
is just a substitution of {\em one} of the $\lambda_L$ by $\lambda_R$. The notation $\lambda_R{\partial\over\partial\lambda_L}$
has the same meaning.

All classes of the ghost number 3 are odd under $\lambda_L\leftrightarrow\lambda_R$.

\paragraph     {Cohomology as a module over the algebra of pure spinors}
The cohomology classes form a module over the commutative algebra
\[
{\cal R}=\mathbb{C}[\lambda_L^1,\ldots, \lambda_L^{16},\;\lambda_R^1,\ldots,\lambda_R^{16}]/\Big((\lambda_L\Gamma^m\lambda_L) = (\lambda_R\Gamma^m\lambda_R) = 0\Big)
\] 
It is generated by: 
\begin{equation}
1,\Psi^{[\lambda\theta]},\Psi^{[\lambda^2\theta^2]}_{scalar},\Psi^{[\lambda^2\theta^2]}_{2-form},\Psi^{[\lambda^3\theta^3]},\Psi^{[\lambda^3\theta^4]},\Psi^{[\lambda^3\theta^5]}
\end{equation}
All other cohomology classes can be obtained from these generators by means
of multiplications of polynormials of $\lambda_L$ and $\lambda_R$. This module, however, is 
not free; its structure is discussed in Sections \ref{Sec:resolution} and \ref{Sec:resolution-compute}.

\subsection{Result in AdS notations}

\subsubsection{Notations and abbreviations}\label{sec:NotationsAndAbbreviations}
\paragraph     {Pure spinors as a cone over the group manifold.}

We find it technically useful to consider two different parametrizations of
spinors. One is the usual picture of 10-dimensional Weyl spinors, which
form a representation of the even part of the Clifford algebra of $so(10)$.
The other is obtained by splitting $so(10)=so(5)\oplus so(5)$ and using
the fact that $so(5)\simeq sp(2)$. From this point of view, each spinor has
an upper and a lower index: $\theta^{\alpha}_a$. Both indices are in the fundamental
representation of the corresponding $so(5)$, and $4\times 4 = 16$. A {\em generic}
pure spinor corresponds to a point of the cone over the group manifold $Sp(2)$;
see Section \ref{sec:GeometryOfPair}.

Therefore there are two different schemes of notations, which we will call
``AdS notations'' and ``flat space notations''.
\begin{center}
\begin{tabular}{|c|c|}
\hline
AdS & flat space
\cr\hline\hline
$\lambda^{[\alpha}_a\omega^{ab}\theta^{\beta]}_b$
&
$(\lambda\Gamma^m\theta)\; \mbox{ \tt\small for }m\in \{0,\ldots 4\}$
\cr\hline
$\lambda^{\alpha}_{[a}\omega_{\alpha\beta}\theta^{\beta}_{b]}$
&
$(\lambda\Gamma^m\theta)\; \mbox{ \tt\small for }m\in \{5,\ldots 9\}$
\cr\hline
 Eq. (\ref{TLT}) & $(\lambda\Gamma^m\theta)\Gamma_m\theta$
\cr\hline
\end{tabular}
\end{center}
The difference between these two schemes is purely notational. There is
no difference between the tangent space at the point of $AdS_5\times S^5$ and
the tangent space at the point of ${\bf R}^{10}$. In particlar, the zero mode BRST
operator (\ref{DefQ0}) is the same in AdS and in flat space.

\paragraph     {Abbreviations}
In the following formulas, we will use following abbreviated notations. We
will write $v^{\bullet}$ instead of $v^{\alpha}$, just to indicate that there is an upper index,
when we don't bother about the value of the index. For example, we write
$\theta^{\bullet}\cap\lambda^{\bullet}$ (or sometimes $(\theta\cap\lambda)^{\bullet\bullet}$) instead of $\theta^{\alpha}_a\omega^{ab}\lambda^{\beta}_b$, just to indicate that
there are two uncontracted indices ($\alpha$ and $\beta$). Similarly, $\theta_{\bullet}\cup\lambda_{\bullet}$ stands for
$\theta^{\alpha}_a\omega_{\alpha\beta}\lambda^{\beta}_b$. We will sometimes omit indices altogether, and simply write $\theta\cup\lambda$.
Furthermore, we will write:
\begin{align}
\{u\cap v\} = \;& u\cap v + v\cap u
\\ 
u^{\bullet}\capzero v^{\bullet} = \;&
u^{\bullet}\cap v^{\bullet} - {1\over 4} \omega^{\bullet\bullet} \;||u\cap v||
\end{align}
Notice that $||u\capzero v|| = 0$ because ${}^{\bullet}\!\cap\cup_{\bullet} = 4\delta^{\bullet}_{\bullet}$.

We will also denote:
\begin{align}\label{TLT}
   [\theta\cap\lambda\cup\theta]_{\Gamma} = \;3\;\theta\cap\lambda\cup\theta +
\left(\lambda{\partial\over\partial\theta}\right)\theta\cap\theta\cup\theta
\end{align}

\subsubsection{Table of cohomology}

\paragraph     {Cohomology of  $\lambda_{L+}{\partial\over\partial\theta_+}$}  This was calculated in \cite{Berkovits:2001rb}; see also
the Appendix of \cite{Krotov:2006th}. In our notations, the nontrivial classes are:
\begin{align}
&   1,
\nonumber\\ 
&\lambda_{L+}\capzero \theta_+ \;,\;\;
\lambda_{L+}\cupzero \theta_+ \;,\;\;
\nonumber\\[5pt]
& \theta_+\cap\lambda_{L+}\cup\theta_+ \;,\;\;
\nonumber\\[2pt]
& \theta_+\cap\lambda_{L+}\cup\theta_+\cap (\lambda_{L+}\cupzero\theta_+)\;,\;\;
\nonumber\\[5pt]
& \theta_+\cup\lambda_{L+}\cap\theta_+\cup\theta_+\cap\lambda_{L+}\cup\theta_+
\;,\;\;
\theta_+\cap\lambda_{L+}\cup\theta_+\cap\theta_+\cup\lambda_{L+}\cap\theta_+
\;,\;\;
\nonumber\\[5pt]
&
||\theta_+\cap\lambda_{L+}\cup\theta_+\cap\theta_+\cup\lambda_{L+}\cap\theta_+
\cup \lambda_{L+}\capzero\theta_+||
\end{align}
In Section \ref{sec:Bicomplex} we will explain how to reduce the calculation of the
cohomology of $Q^{(0)}$ to the cohomology of  $\lambda_{L+}{\partial\over\partial\theta_{+}}$ and  $\lambda_{R+}{\partial\over\partial\theta_{+}}$.

\paragraph     {Cohomology of $Q^{(0)}$}
In Sections \ref{sec:Details} and \ref{sec:FirstQThenD} we will calculate the cohomology of $Q^{(0)}$ up to the
ghost number three. Here we just give the result of the calculation. The
cohomology of $Q^{(0)}$ up to the ghost number three is:
\begin{equation}
 \Phi^{[1]} =  1
\end{equation}
\begin{equation}
   \Phi^{[\lambda]} = (\lambda_{L+} - \lambda_{R+})^{\bullet}_{\bullet}
\end{equation}
\begin{align}
\Phi^{[\lambda\theta]}_{\rm up} = \;&
\{(\lambda_{L+} - \lambda_{R+})\capzero \theta_+\}^{\bullet\bullet}
\\ 
\Phi^{[\lambda\theta]}_{\rm dn} = \;&
\{(\lambda_{L+} - \lambda_{R+})\cupzero \theta_+\}_{\bullet\bullet}
\end{align}
\begin{align}
   \Phi^{[\lambda^2]} = \;& (\lambda_{L+} - \lambda_{R+})^{\bullet}_{\bullet} \;
(\lambda_{L+} - \lambda_{R+})^{\bullet}_{\bullet}
\\ 
\;& \mbox{ \tt\small mod }
\omega^{\bullet\bullet} ((\lambda_{L+} - \lambda_{R+})\cup
(\lambda_{L+} - \lambda_{R+}))_{\bullet\bullet}
\nonumber \\ 
\;& \mbox{ \tt\small and }
\omega_{\bullet\bullet} ((\lambda_{L+} - \lambda_{R+})\cap
(\lambda_{L+} - \lambda_{R+}))^{\bullet\bullet}
\nonumber
\end{align}
\begin{align}
   \Phi^{[\lambda^2\theta]}_{\rm up} = \;&
\{(\lambda_{L+}-\lambda_{R+})\capzero \theta_+\}^{\bullet\bullet}
\; (\lambda_{L+} - \lambda_{R+})^{\bullet}_{\bullet}
\\ 
\;&
\mbox{ \tt\small mod }
(\;\{(\lambda_{L+}-\lambda_{R+})\capzero \theta_+\} \cup
 (\lambda_{L+} - \lambda_{R+})\;)^{\bullet}_{\bullet} \;\; \omega^{\bullet\bullet}
\nonumber \\ 
   \Phi^{[\lambda^2\theta]}_{\rm dn} = \;&
\{(\lambda_{L+}-\lambda_{R+})\cupzero \theta_+\}_{\bullet\bullet}
\; (\lambda_{L+} - \lambda_{R+})^{\bullet}_{\bullet}
\\ 
\;&
\mbox{ \tt\small mod }
(\;\{(\lambda_{L+}-\lambda_{R+})\cupzero \theta_+\} \cap
 (\lambda_{L+} - \lambda_{R+})\;)^{\bullet}_{\bullet} \;\; \omega_{\bullet\bullet}
\nonumber
\end{align}
\begin{align}
   \Phi^{[\lambda^3]} = \;&
(\lambda_{L+} - \lambda_{R+})^{\bullet}_{\bullet}\;\;
(\lambda_{L+} - \lambda_{R+})^{\bullet}_{\bullet}\;\;
(\lambda_{L+} - \lambda_{R+})^{\bullet}_{\bullet}
\\ 
\;&
\mbox{ \small\tt mod some equivalence relations}
\nonumber
\end{align}
\begin{align}
   \Phi^{[\lambda^2\theta^2]}_{\tiny\tt scalar} = \;&
||\lambda_{L+}\cap\theta_+\cup\lambda_{R+}\cap\theta_+||
\\ 
\Phi^{[\lambda^2\theta^2]}_{\tiny\tt up} = \;&
  \lambda_{L+}^{(\bullet}\cap\theta_+\cup\theta_+\cap\lambda_{R+}^{\bullet)}
+  \lambda_{L+}^{(\bullet}\cap\lambda_{R+}\cup\theta_+\cap\theta_+^{\bullet)} \;+
\nonumber \\ 
\;&+\theta_+^{(\bullet}\cap\theta_+\cup\lambda_{L+}\cap\lambda_{R+}^{\bullet)}
+\lambda_{R+}^{(\bullet}\cap\theta_+\cup\lambda_{L+}\cap\theta_+^{\bullet)}\;-
\nonumber \\ 
&
- \theta_+^{(\bullet}\cap\lambda_{R+}\cup\lambda_{L+}\cap\theta_+^{\bullet)}
+ \theta_+^{(\bullet}\cap\lambda_{R+}\cup\theta_+\cap\lambda_{L+}^{\bullet)}
\\ 
\Phi^{[\lambda^2\theta^2]}_{\tiny\tt dn} = \;&
  \lambda_{L+(\bullet}\cup\theta_+\cap\theta_+\cup\lambda_{R+\bullet)}
+  \lambda_{L+(\bullet}\cup\lambda_{R+}\cap\theta_+\cup\theta_{+\bullet)} \;+
\nonumber \\ 
\;&+\theta_{+(\bullet}\cup\theta_+\cap\lambda_{L+}\cup\lambda_{R+\bullet)}
+\lambda_{R+(\bullet}\cup\theta_+\cap\lambda_{L+}\cup\theta_{+\bullet)}\;-
\nonumber \\ 
&
- \theta_{+(\bullet}\cup\lambda_{R+}\cap\lambda_{L+}\cup\theta_{+\bullet)}
+ \theta_{+(\bullet}\cup\lambda_{R+}\cap\theta_+\cup\lambda_{L+\bullet)}
\\ 
\Phi^{[\lambda^2\theta^2]}_{\tiny\tt mixed} = \;&
\Big( \lambda_{R+[\bullet}^{[\bullet} \;
\theta_+^{\bullet]}\cap\lambda_{L+}\cup\theta_{+\bullet]}
\Big)^{\omega-\mbox{\tiny\tt less}}_{\omega-\mbox{\tiny\tt less}} +
\left(\lambda_{L+}{\partial\over\partial \theta_+}\right)\mbox{\tt\small smth}
\end{align}
(Notice that the first term in $\Phi^{[\lambda^2\theta^2]}_{\tiny\tt mixed}$ was called $\stackrel{0}{\Psi}$ in \cite{Mikhailov:2011af}.)

\begin{align}
   \Phi_{\rm up/dn}^{[\lambda^3\theta]} =\;& \Phi^{[\lambda^2\theta]}_{\rm up/dn}
(\lambda_{L+} - \lambda_{R+})^{\bullet}_{\bullet}
\\ 
\;&
\mbox{ \small\tt mod some equivalence relations}
\nonumber
\end{align}
\begin{align}
   \Phi^{[\lambda^3\theta^2]}_{\rm\tiny up/dn/mixed} =\;&
\Phi^{[\lambda^3\theta^2]}_{\rm\tiny up/dn/mixed}(\lambda_{L+} - \lambda_{R+})^{\bullet}_{\bullet}
\\ 
\;&
\mbox{ \small\tt mod some equivalence relations}
\nonumber
\end{align}
\begin{align}
   \Phi^{[\lambda^3\theta^3]}_{{\rm up}} = \;&
\lambda^{(\bullet}_{R+}\cap \theta_+ \cupzero\lambda_{L+}
   \cap \theta_+\cup\lambda_{L+}\cap\theta^{\bullet)}_+ \;+
\nonumber \\ 
+ \; &
\theta^{(\bullet}_+\cap\lambda_{L+}\cup\theta_+\cap\lambda_{L+}\cupzero
\theta_+\cap\lambda^{\bullet)}_{R+}
\; +
\nonumber \\ 
+ \; &
\mbox{\tt\small some }[\lambda_{R+}^2\lambda_{L+}\theta_+^3]
\label{PhiL3T3Up}\\ 
   \Phi^{[\lambda^3\theta^3]}_{{\rm dn}} = \;&
\lambda_{R+(\bullet}\cup \theta_+ \capzero\lambda_{L+}
   \cup \theta_+\cap\lambda_{L+}\cup\theta_{+\bullet)} \;+
\nonumber \\ 
+ \; &
\theta_{+(\bullet}\cup\lambda_{L+}\cap\theta_+\cup\lambda_{L+}\capzero
\theta_+\cup\lambda_{R+\bullet)}
\; +
\nonumber \\ 
+ \; &
\mbox{\tt\small some }[\lambda_{R+}^2\lambda_{L+}\theta_+^3]
\label{PhiL3T3Dn}\\ 
\Phi^{[\lambda^3\theta^3]}_{\rm mixed1} = \;&
\left((\lambda_{R+})_{[\bullet}^{(\bullet}\;
\theta_+^{\bullet)}\capzero\lambda_{L+} \cup \theta_+
\cap\lambda_{L+}\cup \theta_{+\bullet]}\right)_{\omega - \rm less}
\; +
\label{PhiL3T3Mixed1}\\ 
\;& + \mbox{ \tt\small some } [\lambda_{R+}^2\lambda_{L+}\theta_+^3]
\nonumber\\ 
\Phi^{[\lambda^3\theta^3]}_{\rm mixed2} = \;&
\left((\lambda_{R+})_{(\bullet}^{[\bullet}\;
\theta_+^{\bullet]}\capzero\lambda_{L+} \cup \theta_+
\cap\lambda_{L+}\cup \theta_{+\bullet)}\right)^{\omega - \rm less}
\; +
\label{PhiL3T3Mixed2}\\ 
\;& + \mbox{ \tt\small some } [\lambda_{R+}^2\lambda_{L+}\theta_+^3]
\nonumber
\end{align}
\begin{align}
\Phi^{[\lambda^3\theta^4]} =\;& {\partial\over\partial\theta}\;
\Phi^{[\lambda^3\theta^5]}
\label{TypeL3T4}
\end{align}
\begin{align}
&   \Phi^{[\lambda^3\theta^5]} =\;
\nonumber\\
=\;&\left(
1-{5\over 3}\left(\lambda_R{\partial\over\partial\lambda_L}\right)
\right)||
[\theta_+\cap\lambda_{L+}\cup\theta_+]_{\Gamma}\cap
[\theta_+\cup\lambda_{L+}\cap\theta_+]_{\Gamma}\cup
\{\theta\capzero \lambda_{L+}\}
|| -
\nonumber \\ 
& - (\lambda_L\leftrightarrow\lambda_R)
\end{align}
(See the symbolic computation in {\tt L3T5Ansatz}, Section  \ref{sec:Computer}.)

Comments:
\begin{enumerate}
\item Notice that the free indices in $\Phi^{[\lambda^3\theta^3]}_{\rm up}$ and $\Phi^{[\lambda^3\theta^3]}_{\rm dn}$ are symmetrized; similar
expressions with antisymmetrized indices is $Q^{(0)}$-exact
\item We found by a symbolic computation ({\tt L3T5Derivatives} of Section \ref{sec:Computer})
that:
\begin{equation}
\left(\Phi^{[\lambda^3\theta^3]}_{\rm up}\right)^{\alpha\beta} =
\;-{1\over 288}\; \omega^{\alpha\alpha'}\omega^{\beta\beta'}\omega_{ab}
{\partial\over\partial\theta^{\alpha'}_a}{\partial\over\partial\theta^{\beta'}_b}
\Phi^{[\lambda^3\theta^5]}
\end{equation}
This implies that Eq. (\ref{PhiL3T3}) defines a nonzero cohomology class.
\item The $E$-symmetry defines in Eq. (\ref{ESymmetry}) acts as follows:
\begin{equation}
   E\Phi^{[\lambda^3\theta^5]} = -i \Phi^{[\lambda^3\theta^5]}
\end{equation}
This agrees with:
\begin{equation}
   Q \left( \Phi^{[\lambda^2\theta^2]}\; \Omega + \ldots \right) =
   \Phi^{[\lambda^3\theta^5]} \;t^2_mt^2_m\Omega + \ldots
\end{equation}
and $E\Phi^{[\lambda^2\theta^2]} = \Phi^{[\lambda^2\theta^2]}$ --- see (\ref{ETLTM}).
\end{enumerate}

\subsection{Resolution of the cohomology modules}
Notice that the cohomology classes form a module over the commutative algebra
${\cal R}=\mathbb{C}[\lambda_L^1,\ldots, \lambda_L^{16},\;\lambda_R^1,\ldots,\lambda_R^{16}]/\Big((\lambda_L\Gamma^m\lambda_L) = (\lambda_R\Gamma^m\lambda_R) = 0\Big)$. Indeed, we
can multiply any cohomology class by an arbitrary function of $\lambda_L$ and $\lambda_R$, and
get a new cohomology class. But this is not a free module. For example, one 
can see (the computation of {\tt L3T2Eqs} of Section (\ref{sec:Computer})) that
$\lambda_L^{\alpha} \Phi^{[\lambda^2\theta^2]}_{\rm scalar} = 0$ --- this is a {\em relation}. It could be useful to classify
such relations, and also the relations between relations, {\it etc.}

Mathematically, this can be formulated as follows. Let $H^{N,n}$ denote the
cohomology of the ghost number $N$ with the ghost number $N-n$ ({\it i.e.}
with $n$ thetas), and let $\ldots \to M^{(n)}_1 \to M^{(n)}_0\to M^{(n)}\to 0$ denote the minimal
free resolution of the ${\cal R}$-module $M^{(n)}=\bigoplus_{N,n} H^{N,n}$. For every $i$, the free 
module $M_i^{(n)}$ is a tensor product of ${\cal R}$ with some linear space $\mu_i$ on which ${\bf R}$ 
does not act:
\begin{equation}
M^{(n)}_i = \mu^{(n)}_i \otimes_{\bf C} {\cal R}
\end{equation}
Notice that $\mu^{(n)}_i$ is a representation of $Spin(10)\times{\bf C}^{\times}$ --- the group of
rotations and rescaling of $\lambda$ and $\theta_+$. The dimensions of the $so(10)$-modules
$\mu^{(n)}_i$ correspond to the numbers of generators of $M_i$. Having reformulated the
problem in this mathematical language, we observe that it can be solved on a
computer using {\tt Macaulay2} \cite {mac2}.

The information about the free resolution is typically used to find the
structure of the $Spin(10)\times {\bf C}^{\times}$-module on $\mu^{(n)}_i$ and therefore on $M^{(n)}$. We 
can proceed in the opposite direction, using the information about the 
structure of the $Spin(10)\times {\bf C}^{\times}$ -module on $M$ to find the structure of $\mu_i$ 
using the formula:
$$\sum_i(-1)^i\mu_i \otimes\mathcal{S}^2=\sum _N H^N\tau^N ,$$
where ${\cal S}$ is a formal linear combination of $so(10)$ weights:
\begin{align}
   {\cal S} =\;& \sum_{m=0}^{\infty} a_m\tau^m
\\ 
a_m =\;& [0,0,0,0,m]
\end{align}
Let us define ${\cal S}^{-2} = \sum^\infty_{n=0}b_n\tau^n$, so that it satisfies:
$$({\cal S}^2)({\cal S}^{-2})=
(\displaystyle\sum^\infty_{m=0}a_m\tau^m)^{\otimes 2}\otimes
(b_0+b_1\tau+\cdots+b_i\tau^i+\cdots)=1$$
We have $b_0=1$, $b_1= -2 a_1$, {\it etc.} Then we get:
\begin{equation}
 \sum_i(-1)^i\mu _i =\sum _N H^N\tau^N \otimes \mathcal{S}^{-2}, \label{Eq:alt_mu}
\end{equation}
The analysis of the resolution of the cohomology module is given in the Sec.~\ref{Sec:resolution-compute}.

\label{Sec:resolution}
\section{Details of calculation}\label{sec:Details}
\subsection{Some vanishing theorems which follow immediately}
The following classes are necessarily zero:
\begin{align}
   \lambda\theta^{\geq 3}\;,\;\;\lambda^2\theta^{\geq 5}\;,\;\;\lambda^3\theta^{\geq 6}
\end{align}
as follows from considering the term with the maximal power of $\lambda_L$.

\subsection{Spectral sequence of a bicomplex}\label{sec:Bicomplex}
One method to compute the cohomology of  $Q^{(0)}$ uses the spectral sequence
of some bicomplex, which we will now describe.

\subsubsection{Bicomplex}
\paragraph     {Introducing $\sigma$ and $d$}
We will consider the spectral sequence corresponding to the
following two differentials:
\begin{align}
Q = &\; Q_L + Q_R
\\ 
&\;\mbox{\tt\small where }
Q_L =  \lambda^{\alpha}_L{\partial\over\partial\theta^{\alpha}_L}
\;\;\mbox{ \tt\small and }\;\;
Q_R = \lambda^{\alpha}_R{\partial\over\partial\theta^{\alpha}_R}
\\ 
d = &\; \sigma^{\alpha}\left({\partial\over\partial\theta^{\alpha}_L} - {\partial\over\partial\theta^{\alpha}_R}\right)
\end{align}
Here $\sigma^{\alpha}$ is a new bosonic variable.

\paragraph     {$\bf Z$ - grading}
All these differentials respect the total degree $N$:
\begin{align}
   N = \theta_L{\partial\over\partial\theta_L}
   + \theta_R{\partial\over\partial\theta_R}
   + \lambda_L{\partial\over\partial\lambda_L}
   + \lambda_R {\partial\over\partial\lambda_R}
   + \sigma{\partial\over\partial\sigma}
\end{align}
\paragraph     {Symmetry ${\bf Z}_2^{LR}$}
Notice that both $Q$ and $d$ are invariant under the ${\bf Z}_2$ symmetry
which exchanges $L\leftrightarrow R$ and $\sigma\to -\sigma$. We will call it ${\bf Z}_2^{LR}$.

\paragraph     {Total complex}
We introduce the total complex, with the differential:
\begin{equation}
   Q_{\rm tot} = Q + d
\end{equation}
It turns out that our problem is equivalent to calculating the cohomology of
$Q_{\rm tot}$. To prove this, we have to remember how the cohomology of the bicomplex
is calculated using the spectral sequences.

\subsubsection{Spectral sequence}
\paragraph     {First $d$ then $Q$}
The first method is to first calculate the cohomology of $d$, and then
consider $Q$ as a small perturbation. The cohomology of $d$ is:
\begin{equation}
   H(d) = \mbox{Fun}(\;\theta_L + \theta_R \;,\; \lambda_L \;,\; \lambda_R\;)
\end{equation}
This gives the ``first page'' of the spectral sequence $\widetilde{E}$, {\it i.e.} $\widetilde{E}_1$.
Notice that everything is graded by $N$. We get:
\begin{align}
   \widetilde{E}_1^{p>0,q}[N] = &\; 0
\\ 
\widetilde{E}_1^{0,q}[N] = &\; [\lambda_{L,R}^q (\theta_L+\theta_R)^{N-q}]
\end{align}
This implies that this spectral sequence terminates on the first page.
Therefore:
\begin{itemize}
\item the cohomology of $Q+d$ is equal to the cohomology of $Q$ on expressions
   which depend on $\theta_{L,R}$ only in the combination $\theta_L + \theta_R$
\end{itemize}
This is exactly what we want to calculate.

\paragraph     {First $Q$ then $d$}
The idea is to calculate first the cohomology of $Q$, and then act by $d$ on it.
We will develop this idea in Section \ref{sec:FirstQThenD}.

\subsection{Symbolic computations using {\tt Macaulay2} and {\tt LiE}}\label{Sec:M2Lie}
Another method is to use the symbolic computations. There are several tools
which we will describe in this and the following section.

\subsubsection{Computation of cohomology}
\paragraph     {Notations and setup}
Here we will use the description of the representations of $\sso (10)$ using
weight diagramms:
\begin{align}
   [1,0,0,0,0] \quad & \mbox{\small\tt vector}
\\ 
[0,1,0,0,0] \quad & \mbox{\small\tt antisymmetric 2-form}
\\ 
[0,0,1,0,0] \quad & \mbox{\small\tt antisymmetric 3-form}
\\ 
[0,0,0,1,0] \quad & \mbox{\small\tt antichiral spinor}
\\ 
[0,0,0,0,1] \quad & \mbox{\small\tt chiral spinor}
\end{align}
We want to calculate the cohomology of the differential $Q=(\lambda_L+\lambda_R)\frac{\partial}{\partial \theta}$
where $\theta$ is an odd ten-dimensional spinor transforming according the
representation  $[0,0,0,0,1]$ of  $\sso (10)$ and $\lambda_L,\lambda_R$ are pure spinors
transforming according the same representation. {\footnote { As usual the representations are labeled by coordinates of their highest weight. The vector representation $V$ has the highest weight $[1,0,0,0,0]$, the irreducible spinor representations have highest weights $[0,0,0,0,1]$, $[0,0,0,1,0]$.} More details about the
computation procedure in this section could be referred to \cite{Movshev:2011pr}}. 

We will describe these cohomology groups as representations of the Lie
algebra $\sso{(10)}$.
Our calculations in this section use the computer programs {\tt Macaulay2}\cite{mac2}
and {\tt LiE} \cite{LiEcode}. More precisely, we consider the differential $Q$ acting on
chain complex with components
$$\displaystyle\sum^\infty_{m_L=0}[0,0,0,0,m_L] \otimes \displaystyle\sum^\infty_{m_R=0}[0,0,0,0,m_R] \otimes \Lambda^n{[0,0,0,0,1]}.$$
Here $[0,0,0,0,m_L]$ can be identified with the space of polynomial functions
of the pure spinor $\lambda_L$ of the order $m_L$, and $[0,0,0,0,m_R]$ with the space
of polynomial functions of the pure spinor $\lambda_R$ of the order $m_R$.

We apply the {\tt LiE} program to obtain the decomposition of this complex
into irreducible representations and use the dimensions of cohomology found
by  means of \cite{mac2} to describe the action of the differential. The package
{\tt DGAlgebras} of {\tt Macaulay2} already has procedures for
calculating the cohomologies of the Koszul complex.

\paragraph     {Results of computations}
The cohomology group has two gradings:  $N=m_L+m_R+n$ and $n$:
\begin{equation}
   H = \bigoplus_{N,n} H^{N,n}
\end{equation}
Using LiE, we could explicitly describe the graded components of the
cohomology group, $H^{N,n}$, with gradings by the following general formulas
valid for $N\neq 4$:
\beqr
H^{N,0}&=& [0,0,0,0,N] \label{Hk,0}\\
H^{N,1}&=& [1,0,0,0,N-2] \\
H^{N,2}&=& [0,1,0,0,N-4] \\
H^{N,3}&=& [0,0,1,0,N-6] \\
H^{N,4}&=& [0,0,0,1,N-7] \\
H^{N,5}&=& [0,0,0,0,N-8] \label{Hk,5}
\eeqr
When $N=4$, there is one additional term, a scalar, in $H^{4,2}$:
\beq
H^{4,2}=[0,0,0,0,0] \oplus [0,1,0,0,0] \label{H4,2}
\eeq
The $\SO(10)$-invariant part is in $H^{0,0}$, $H^{8,5}$, and $H^{4,2}$.

The dimensions of these cohomology groups are encoded in series
$P_n(\tau)=\sum_N \dim H^{N,n}\tau^N$ (Poincar\'e series) that can be calculated by means
of {\tt Macaulay2}~\cite{mac2}:
\begin{eqnarray*}
P_0(\tau)&=&\frac{1 + 5 {\tau} + 5 {\tau}^2 + {\tau}^3}{(1 - {\tau})^{11}}\\
P_1(\tau)&=&\frac{10 {\tau}^2 + 34 {\tau}^3 + 16 {\tau}^4)}{(1 - {\tau})^{11}}\\
P_2(\tau)&=&(46 {\tau}^4 + 54 {\tau}^5 + 66 {\tau}^6 - 166 {\tau}^7 + 330 {\tau}^8 - 462 {\tau}^9 + 462 {\tau}^{10} - 330 {\tau}^{11} \nonumber\\
 && + 165 {\tau}^{12} - 55 {\tau}^{13} + 11 {\tau}^{14} - {\tau}^{15})/{(1 - {\tau})^{11}}\\
P_3(\tau)&=&(120 {\tau}^6 - 120 {\tau}^7 + 330 {\tau}^8 - 462 {\tau}^9 + 462 {\tau}^{10} - 330 {\tau}^{11} + 165 {\tau}^{12} \nonumber\\
 && - 55 {\tau}^{13} + 11 {\tau}^{14} - {\tau}^{15})/{(1 - {\tau})^{11}}\\
P_4(\tau)&=&\frac{16 {\tau}^7 + 34 {\tau}^8 + 10 {\tau}^9}{(1 - {\tau})^{11}}\\
P_5(\tau)&=&\frac{{\tau}^8 + 5 {\tau}^9 + 5 {\tau}^{10} + {\tau}^{11}}{(1 - {\tau})^{11}}\\
\end{eqnarray*}
The cohomology $H^n=\bigoplus_N H^{N,n}$ can be regarded as a
${\C}[\lambda_L^1,\ldots, \lambda_L^{16},\;\lambda_R^1,\ldots,\lambda_R^{16}]$-module. Using {\tt Macaulay2} one can obtain the
number of its generators. The number of $0,\cdots,5$-th cohomology generators are
$1,10,46,120,16,1$, respectively. Using the highest weight vector
representation, beside of $1$, they are
\beqr
H^{2,1}&=&[1,0,0,0,0], \\
H^{4,2}&=&[0,0,0,0,0] +[0,1,0,0,0],\\
H^{6,3}&=&[0,0,1,0,0],\\
H^{7,4}&=&[0,0,0,1,0],\\
H^{8,5}&=&[0,0,0,0,0].
\eeqr
The expressions for generators are given in Sec.\ref{sec:FlatSpaceNotations} , specifically:
\begin{center}
\begin{tabular}{|c|c|c|}
\hline
$H^{2,1}$ & $[1,0,0,0,0]$ & $\Phi^{[\lambda\theta]}$ \\   
\hline
$H^{4,2}$ & $[0,0,0,0,0]$ & $\Phi^{[\lambda^2\theta^2]}_{\rm scalar}$ \\  
      & $[0,1,0,0,0]$ & $\Phi^{[\lambda^2\theta^2]}_{\rm 2-form}$ \\  
\hline
$H^{6,3}$ & $[0,0,1,0,0]$ & $\Phi^{[\lambda^3\theta^3]}$ \\   
\hline
$H^{7,4}$ & $[0,0,0,1,0]$ & $\Phi^{[\lambda^3\theta^4]}$ \\  
\hline
$H^{8,5}$ & $[0,0,0,0,0]$ & $\Phi^{[\lambda^3\theta^5]}$ \\
\hline
\end{tabular}
\end{center}

\paragraph     {Generating cohomology by differentiation with respect to $\theta$}
Some generators can be obtained from the generator of $H^{8,5}$ (denoted later by
$\Psi$) by means of differentiation with respect to $\theta$.
Namely, the generators belonging to $H^{7,4}$ are equal to $\frac{\partial \Psi}{\partial \theta^{\alpha}}$, the generators
belonging to $H^{6,3}$ are equal to $\Gamma'^{\alpha \beta}_{abc}\frac{\partial^2}{\partial \theta^{\alpha}\partial \theta ^{\beta}}\Psi,$ where $\Gamma'$ is some matrix
anti-symmetric with respect to $a,b,c$, and not symmetric with respect to $\alpha,\beta$.
One can find the minimal number of generators having the property that all
other generators can be  obtained from them by means of differentiation with
respect to $\theta.$
To calculate this number  we notice, that the cohomology can be considered
also as module over the ring ${\C}[\lambda_L^1,\ldots, \lambda_L^{16},\;
\lambda_R^1,\ldots, \lambda_R^{16}]\otimes \Lambda[b]$ where $b_{\alpha}$
stands for $\frac{\partial}{\partial\theta^{\alpha}}$. Calculations with {\tt Macaulay2}  allow us to calculate the
number of generators of this module.

It is equal to $58$. This means that the module we are interested in is
generated by $1, H^{2,1}=[1,0,0,0,0],
H^{4,2}=[0,0,0,0,0] +[0,1,0,0,0],$
and $H^{8,5}=[0,0,0,0,0].$

\paragraph     {Behaviour under the exchange $\lambda_L\leftrightarrow\lambda_R$}
Notice that our differential is invariant with respect to the involution
$\lambda_L\to\lambda_R.$ Therefore this involution acts on homology. The cohomology groups
$H^{k,0}$ and $H^{k,2}$ (except  $[0,0,0,0,0]$ in $H^{4,2}$) are invariant  (even)  with
respect to this involution. Other cohomology groups are odd ({\it i.e.} the
involution acts as multiplication by $-1$). In particular, the generators $1$
and $[0,1,0,0,0]$ in $H^{4,2}$ are even, other generators are odd.

\subsubsection{Resolution of the cohomology modules}\label{Sec:resolution-compute}

Based on the method discussed in Sec. \ref{Sec:resolution}, one can find a minimal free resolution of the $\mathcal{R}$-module $M=\sum _N H^{N,n}$, where $\mathcal{R}=\mathbb{C}[\lambda_L^1,\ldots, \lambda_L^{16},\;\lambda_R^1,\ldots,\lambda_R^{16}]=\displaystyle\sum^\infty_{m_L=0}[0,0,0,0,m_L]\otimes \displaystyle\sum^\infty_{m_R=0}[0,0,0,0,m_R].$
 The reader may wish to consult \cite{CE} on this subject. The free resolution has the form
$$\cdots \to M_i\to \cdots \to M_0\to M\to 0$$
where $M_i=\mu_i\otimes \mathcal{R},$ and \\
$\mu_0$ - generators of $M$;\\
$\mu_1$ - relations between generators of $M$;\\
$\mu_2$ - relations between relations ;\\
$\cdots$

We give the structure of $\mu_i$ as $so(10)$-module.
\begin{itemize}
\item $n=0,$
$$\mu_0=[0,0,0,0,0], \dim(\mu_0)=1, \deg(\mu_0)=0;$$
$$\mu_1=[0,0,0,0,1], \dim(\mu_1)=16, \deg(\mu_1)=1;$$
$$\mu_2=[0,0,1,0,0]+[1,0,0,0,0], \dim(\mu_2)=130, \deg(\mu_2)=2;$$
$$\mu_3=2\times[0,0,0,1,0] +[0,1,0,1,0] +[1,0,0,0,1], \dim(\mu_3)=736, \deg(\mu_3)=3;$$
\begin{equation*}
  \begin{aligned}
\mu_4=&2\times[0,0,0,0,0] +2\times[0,0,0,1,1] +3\times[0,1,0,0,0] +[0,2,0,0,0] +\\
&+[1,0,0,2,0] +[1,0,1,0,0] +[2,0,0,0,0], \dim(\mu_4)=3376, \deg(\mu_4)=4;
\end{aligned}
\end{equation*}
$$\cdots$$

\item $n=1,$
$$\mu_0=[1,0,0,0,0], \dim(\mu_0)=10, \deg(\mu_0)=2; $$
$$\mu_1=2\times[0,0,0,1,0] +[1,0,0,0,1], \dim(\mu_1)=176, \deg(\mu_1)=3; $$
\begin{equation*}
    \begin{aligned}
\mu_2=&3\times[0,0,0,0,0] +2\times[0,0,0,1,1] +4\times[0,1,0,0,0] +[1,0,1,0,0]+ \\ 
&+[2,0,0,0,0], \dim(\mu_2)=1602, \deg(\mu_2)=4;
    \end{aligned}
\end{equation*}
\begin{equation*}
    \begin{aligned}
\mu_3=&8\times[0,0,0,0,1] +2\times[0,0,1,1,0] +4\times[0,1,0,0,1] +8\times[1,0,0,1,0] + \\
&+[1,1,0,1,0] +[2,0,0,0,1], \dim(\mu_3)=10336, \deg(\mu_3)=5;
    \end{aligned}
\end{equation*}
$$\cdots$$

\item $n=2,$
$$\mu_0=[0,0,0,0,0] +[0,1,0,0,0], \dim(\mu_0)=46, \deg(\mu_0)=4; $$
$$\mu_1=4\times[0,0,0,0,1] +[0,1,0,0,1] +2\times[1,0,0,1,0], \dim(\mu_1)=912, \deg(\mu_1)=5; $$
\begin{equation*}
    \begin{aligned}
\mu_2=&3\times[0,0,0,0,2] +3\times[0,0,0,2,0] +8\times[0,0,1,0,0]+\\
&+[0,1,1,0,0] +9\times[1,0,0,0,0] +2\times[1,0,0,1,1]+\\
&+4\times[1,1,0,0,0], \dim(\mu_2)=9512, \deg(\mu_2)=6;
    \end{aligned}
\end{equation*}
$$\cdots$$

\item $n=3,$
$$\mu_0=[0,0,1,0,0], \dim(\mu_0)=120, \deg(\mu_0)=6; $$
\begin{equation*}
    \begin{aligned}
\mu_1=&2\times[0,0,0,1,0] +[0,0,1,0,1] +2\times[0,1,0,1,0] +2\times[1,0,0,0,1], \\
&\dim(\mu_1)=2640, \deg(\mu_1)=7;
    \end{aligned}
\end{equation*}
\begin{equation*}
    \begin{aligned}
\mu_2=&3\times[0,0,0,0,0] +8\times[0,0,0,1,1] +[0,0,2,0,0] +7\times[0,1,0,0,0] +\\
&+2\times[0,1,0,1,1] +3\times[0,2,0,0,0] +2\times[1,0,0,0,2] +3\times[1,0,0,2,0] +\\
&+5\times[1,0,1,0,0] +3\times[2,0,0,0,0], \dim(\mu_2)=30450, \deg(\mu_2)=8;
    \end{aligned}
\end{equation*}
$$\cdots$$

\item $n=4,$
$$\mu_0=[0,0,0,1,0], \dim(\mu_0)=16, \deg(\mu_0)=7; $$
$$\mu_1=2\times[0,0,0,0,0] +[0,0,0,1,1] +2\times[0,1,0,0,0], \dim(\mu_1)=302, \deg(\mu_1)=8; $$
\begin{equation*}
    \begin{aligned}
\mu_2=&5\times[0,0,0,0,1] +[0,0,1,1,0] +2\times[0,1,0,0,1] +4\times[1,0,0,1,0],\\
& \dim(\mu_2)=2976, \deg(\mu_2)=9;
    \end{aligned}
\end{equation*}
\begin{equation*}
    \begin{aligned}
\mu_3=&3\times[0,0,0,0,2] +6\times[0,0,0,2,0] +10\times[0,0,1,0,0] +[0,1,0,2,0] +\\
&+2\times[0,1,1,0,0] +10\times[1,0,0,0,0] +4\times[1,0,0,1,1] +\\
&+6\times[1,1,0,0,0], \dim(\mu_3)=20902, \deg(\mu_3)=10;
    \end{aligned}
\end{equation*}
\begin{equation*}
    \begin{aligned}
\mu_4=&24\times[0,0,0,1,0] +6\times[0,0,0,2,1] +8\times[0,0,1,0,1] +\\
&+23\times[0,1,0,1,0]+2\times[0,2,0,1,0] +24\times[1,0,0,0,1] +\\
&+[1,0,0,3,0] +4\times[1,0,1,1,0] +6\times[1,1,0,0,1] +\\
&+9\times[2,0,0,1,0], \dim(\mu_4)=120224, \deg(\mu_4)=11;
    \end{aligned}
\end{equation*}
$$\cdots$$

\item $n=5,$
$$\mu_0=[0,0,0,0,0], \dim(\mu_0)=1, \deg(\mu_0)=8; $$
$$\mu_1=[0,0,0,0,1], \dim(\mu_1)=16, \deg(\mu_1)=9; $$
$$\mu_2=[0,0,1,0,0]+[1,0,0,0,0], \dim(\mu_2)=130, \deg(\mu_2)=10; $$
$$\mu_3=2\times[0,0,0,1,0] +[0,1,0,1,0] +[1,0,0,0,1], \dim(\mu_3)=736, \deg(\mu_3)=11; $$
\begin{equation*}
  \begin{aligned}
\mu_4=&2\times[0,0,0,0,0] +2\times[0,0,0,1,1] +3\times[0,1,0,0,0] +[0,2,0,0,0] +\\
&+[1,0,0,2,0] +[1,0,1,0,0] +[2,0,0,0,0], \dim(\mu_4)=3376, \deg(\mu_4)=12;
    \end{aligned}
\end{equation*}
$$\cdots$$
 where $i\times[a,b,c,d,e]$ denotes the representation $[a,b,c,d,e]$ with multiplicity $i$.

\end{itemize}

\subsection{Symbolic computations using a canonical form of a pair of pure spinors}
\subsubsection{Geometry of a pair of pure spinors in $AdS_5\times S^5$}\label{sec:GeometryOfPair}
As was explained in \cite{Mikhailov:2011af}, the orbit of the generic pure spinor under
$so(5)\oplus so(5) = sp(2)\oplus sp(2)$ is a cone over
the group manifold of the symplectic group $Sp(2)$:
\begin{align}
   (\lambda_{R+})^{\alpha}_a\omega^{ab}(\lambda_{R+})^{\beta}_b \;= \;&
{1\over 4}||\lambda_{R+}\cap\lambda_{R+}||\; \omega^{\alpha\beta}
\\ 
   (\lambda_{R+})^{\alpha}_a\omega_{\alpha\beta}(\lambda_{R+})^{\beta}_b \;= \;&
{1\over 4}||\lambda_{R+}\cap\lambda_{R+}||\; \omega_{ab}
\end{align}
(These two equations are equivalent when $||\lambda_{R+}\cap\lambda_{R+}||\neq 0$.) The same
constraints are imposed on $\lambda_{L+}$.

Let us consider the orbit with $||\lambda_{R+}\cap\lambda_{R+}|| = 4c_R$ where $c_{R}$ is a complex
number. In this case we can use the change of variables:
\begin{equation}
   (\lambda_{R+})^{\alpha}_a \mapsto M_a^b(\lambda_{R+})^{\alpha}_b
\end{equation}
where $M\in Sp(2)$, to ``diagonalize'' $\lambda_{R+}$:
\begin{equation}\label{CanonicalLambdaR}
 (\lambda_{R+})^{\alpha}_a = c_R\delta^{\alpha}_a
\end{equation}
This choice of $\lambda_{R+}$ breaks $sp(2)\oplus sp(2)$ to a diagonal $sp(2)$.

In a generic case, this residual $sp(2)$ can be used to bring $\lambda_{L+}$ to the
form:
\begin{equation}\label{CanonicalLambdaL}
   \lambda_{L+} = c_L\left(\begin{array}{cccc}
         a & 0     & 0 & 0     \cr
         0 & a^{-1} & 0 & 0     \cr
         0 & 0     & b & 0     \cr
         0 & 0     & 0 & b^{-1}
      \end{array}\right)
\end{equation}
This is only true when $\lambda_{L+}$ is in generic position with respect to $\lambda_{R+}$.
This is because a {\em generic} quadratic Hamiltonian $H(q_1,p_1,q_2,p_2)$ can be
brought to the form $\alpha q_1p_1 + \beta q_2p_2$ by a canonical transformation\footnote{in this particular calculation we do not care about reality}.

We used this explicit parametrization of the pair of pure spinors in our
computer calculation: see Section \ref{sec:Computer}.

We conclude that there is a $4$-parameter family of the orbits of maximal
dimension $22 - 4 = 18$. Notice that the dimension of $sp(2)\oplus sp(2)$ is $20$,
therefore there should be $20-18=2$ elements of the diagonal part of
$sp(2)\oplus sp(2)$ stabilizing the pair $\lambda_{L+}, \lambda_{R+}$.
They are parametrized by $\nu_1$ and $\nu_2$:
\begin{equation}
   \left(\begin{array}{cccc}
         \nu_1 & 0     & 0 & 0     \cr
         0 & -\nu_1 & 0 & 0     \cr
         0 & 0     & \nu_2 & 0     \cr
         0 & 0     & 0 & -\nu_2
      \end{array}\right) \;\in \; sp(2)
\end{equation}
Notice under that the action of $Spin(10)\times {\bf C}^{\times}\times {\bf C}^{\times}$, any two {\em generic}
pairs $(\lambda^{(1)}_L,\lambda^{(1)}_R)$ and $(\lambda^{(2)}_L,\lambda^{(2)}_R)$ are equivalent, {\it i.e.} there are no
invariants. But if we restrict to $so(5)\oplus so(5)$, then there are invariants
$c_L,c_R,a,b$.

\subsubsection{Computer program for direct calculation of cohomology}\label{sec:Computer}
We have a computer program which does symbolic manipulations with
elements of free supercommutative algebras. This allows a straightforward
symbolic computation of the BRST cohomology. The program is available here:

{\tt http://code.google.com/p/minitheta/w/list}

\vspace{15pt}
\noindent
The idea is to straightforwardly compute the cohomology of the operator
$(\lambda_{L+} + \lambda_{R+}){\partial\over\partial\theta_+}$ on expressions polynomial in $\lambda_{L+},\lambda_{R+},\theta_+$. We write all the
possible $sp(2)\oplus sp(2)$-covariant polynomials which can be constructed from
$\lambda_{L+},\lambda_{R+},\theta_+$ and the symplectic forms $\omega_{\alpha\beta}$ and $\omega^{ab}$, and then compute the
action of $(\lambda_{L+} + \lambda_{R+}){\partial\over\partial\theta_+}$ on the space of these polynomials.
The pure spinor constraints on $\lambda_{L+}$ and $\lambda_{R+}$ are taken into account by
subsitution of expressions (\ref{CanonicalLambdaL}) and (\ref{CanonicalLambdaR}) for $\lambda_{L+}$ and $\lambda_{R+}$.

\section{Use of spectral sequence}\label{sec:FirstQThenD}
In this section we will compute the cohomology using the spectral sequence
for the bicomplex $Q + d$ of Section \ref{sec:Bicomplex}. We have seen in Section \ref{sec:Bicomplex} that the
cohomology of $Q+d$ is the same as the cohomology of our $Q^{(0)}$. Here we will
calculate this cohomology by first computing the cohomology of $Q$ (which is
well known) and then treating $d$ as a perturbation.

\vspace{15pt}
\noindent
The cohomology of $Q$ is well known:
\begin{align}
   H(Q) = \mbox{Fun}\Big( \sigma,\; &
   [\lambda_L\theta_L],\;[\lambda_L\theta_L^2],\;[\lambda_L^2\theta_L^3],\;
   [\lambda_L^2\theta_L^4],\;[\lambda_L^3\theta_L^5],
   \nonumber \\ 
   &
   [\lambda_R\theta_R],\;[\lambda_R\theta_R^2],\;[\lambda_R^2\theta_R^3],\;
   [\lambda_R^2\theta_R^4],\;[\lambda_R^3\theta_R^5]\;
   \Big)\;/\;\mbox{Im}(Q)
\end{align}
We identify:
\begin{align}
   E_1^{p,q}[N] = \mbox{ \small\tt elements of the form }
[\sigma^p \lambda^q \theta^{N-p-q}] \mbox{ \small\tt in } H(Q)
\end{align}
As we explained in Section \ref{sec:Bicomplex}, $N$ commutes with both $Q$ and $d$.

\subsection{Case $N=0$}
We have $H(Q)[0]={\bf C}$ --- just constants. In this case $d$ acts trivially
and we get:
\begin{equation}
   H(d+Q)[0] = {\bf C}
\end{equation}

\subsection{Case $N=1$}
We observe that $H(Q)[1]$ is generated by $\sigma$:
\begin{align}
   E_1^{1,0}[1] = &\; {\bf C}\mbox{ \tt \small generated by }\sigma
\\ 
E_1^{0,1}[1] = &\; 0
\end{align}
and all other $E_1^{p,q} = 0$. Therefore $E^1_{\infty} = \widetilde{E}^1_{\infty} = {\bf C}$, in fact $\widetilde{E}^1_{\infty}$ is generated
by $\lambda_L - \lambda_R$. Indeed, $\lambda_L - \lambda_R$ is $Q$-closed and cannot be obtained as
$Q$ of an expression involving $\theta_L$ and $\theta_R$ only through $\theta_L + \theta_R$. Therefore, it
represents a nontrivial cohomology class:
\begin{align}
   H(d+Q)[1] =  {\bf C}\;& \mbox{ \tt \small generated by either }
\sigma \mbox{ \tt \small or }
\lambda_L - \lambda_R
\nonumber \\
 &
\mbox{ \tt\small depending on the point of view}
\end{align}

\subsection{Case $N=2$}
Notice that $H(Q)[2]$ is generated by the following elements:
\begin{align}
E_1^{0,0}[2] = E_1^{1,0}[2] = &\;0
\\ 
 E_1^{0,1}[2] = &\; V\oplus V \mbox{ \tt\small generated by }
(\lambda_L\Gamma^m\theta_L) \mbox{ \tt\small and } (\lambda_R\Gamma^m\theta_R)
\\ 
   E^{2,0}_1[2] = &\;
   S^2{\cal S} \mbox{ \tt\small generated by } \sigma^{\alpha}\sigma^{\beta}
\\ 
E^{1,1}_1[2] = &\;0
\\ 
E^{0,2}_1[2] = &\;0
\end{align}
We observe that $d_1: E_1^{p,q}[2] \to E_1^{p+1,q}[2]$ is zero, therefore $E^{p,q}_2[2] = E^{p,q}_1[2]$.
However, the $d_2$ acts nontrivially:
\begin{align}
   d_2\;:\;E_2^{0,1}[2] \to&\; E_2^{2,0}[2]
\\ 
d_2\left( (\lambda_L\Gamma^m\theta_L) + (\lambda_R\Gamma^m\theta_R) \right)
= &\; (\sigma\Gamma^m\sigma)
\\ 
d_2\left( (\lambda_L\Gamma^m\theta_L) - (\lambda_R\Gamma^m\theta_R) \right)
= &\; 0
\end{align}
We conclude that $d_2$ cancels $(\lambda_L\Gamma^m\theta_L) + (\lambda_R\Gamma^m\theta_R)$ against $(\sigma\Gamma^m\sigma)$ and
we are left with the following:
\begin{itemize}
   \item $E_{\infty}[2]$ is generated by $((\lambda_L -\lambda_R)\Gamma^m (\theta_L + \theta_R))$ and $(\sigma\Gamma^{m_1\ldots m_5}\sigma)$
   \item we also observe that $(\sigma \Gamma^{m_1\ldots m_5}\sigma)$ is $d+Q$-equivalent to:
\begin{equation}
   \left(
      (\lambda_L - \lambda_R)\Gamma^{m_1\ldots m_5}(\lambda_L - \lambda_R)
   \right)
\end{equation}
\end{itemize}
The expression $   \left(
      (\lambda_L - \lambda_R)\Gamma^{m_1\ldots m_5}(\lambda_L - \lambda_R)
   \right)$ is in cohomology, in a sense
that it cannot be obtained as $Q$ of something which only depends on $\theta_L + \theta_R$.

\subsection{Case $N=3$}
\subsubsection{$E^{p,q}_1[3]$}
$H(Q)[3]$ is generated by:
\begin{align}
   E_1^{0,0}[3] = E_1^{1,0}[3] = &\; 0
\\ 
E_1^{0,1}[3] = &\; {\cal S}'\oplus {\cal S}'
\mbox{ \tt\small generated by } \lambda_L\theta_L^2 \mbox{ and }
\lambda_R\theta_R^2
\\ 
E_1^{0,2}[3] = E_1^{2,0}[3] = &\; 0
\\ 
E_1^{1,1}[3] = &\; {\cal S}\otimes (V\oplus V)\mbox{ \tt\small generated by }
\nonumber \\ 
&\;
\sigma^{\alpha}(\theta_L\Gamma^m\lambda_L) \mbox{ \tt\small and }
\sigma^{\alpha}(\theta_R\Gamma^m\lambda_R)
\\ 
E_1^{3,0}[3] = &\; S^3{\cal S}
\mbox{ \tt\small generated by } \sigma^{\alpha}\sigma^{\beta}\sigma^{\gamma}
\\ 
E_1^{2,1}[3] = E_1^{1,2}[3] = E_1^{0,3}[3] = &\; 0
\end{align}
The differential $d_1: E_1^{p,q}[3] \to E_1^{p+1,q}[3]$ acts nontrivially in the following
components:
\begin{align}
   d_1\;:\;E_1^{0,1}[3] \rightarrow \;& E_1^{1,1}[3]
\nonumber \\ 
d_1\left((\lambda_L\Gamma_m\theta_L)\Gamma^m\theta_L\right)\simeq \;&
(\lambda_L\Gamma_m\theta_L)\Gamma^m\sigma
\label{D1OnE1013L}
\\ 
d_1\left((\lambda_R\Gamma_m\theta_R)\Gamma^m\theta_R\right)\simeq \;&
(\lambda_R\Gamma_m\theta_R)\Gamma^m\sigma
\label{D1OnE1013R}
\end{align}
\subsubsection{$E_2^{p,q}[3]$}
Eqs. (\ref{D1OnE1013L}) and (\ref{D1OnE1013R}) imply that:
\begin{align}
E_2^{0,1}[3]=&\;0
\\ 
E_2^{1,1}[3]=&\;
({\cal S}\otimes (V\oplus V))/({\cal S}'\oplus {\cal S}')
\mbox{ \tt\small where } {\cal S}'\oplus {\cal S}'
\nonumber \\ 
&\;
\mbox{ \tt\small is generated by }
(\lambda_L\Gamma_m\theta_L)\Gamma^m\sigma \mbox{ \tt\small and }
(\lambda_R\Gamma_m\theta_R)\Gamma^m\sigma
\end{align}
and the other components of $E_2[3]$ are the same as the corresponding
components of $E_1[3]$. In particular, $E_2^{3,0}[3] = E_1^{3,0}[3] = S^3{\cal S}$.

\subsubsection{$E_3^{p,q}[3]$}
There is a nontrivial $d_2$:
\begin{align}
d_2\;:\;E_2^{1,1}[3]\rightarrow \;& E_2^{3,0}[3]
\\ 
d_2\;\left( \sigma^{\alpha}(\theta_L\Gamma^m\lambda_L) \right) =\;&
\sigma^{\alpha}(\sigma\Gamma^m\sigma)
\\ 
d_2\;\left( \sigma^{\alpha}(\theta_R\Gamma^m\lambda_R) \right) =\;&
\sigma^{\alpha}(\sigma\Gamma^m\sigma)
\end{align}
(The overall coefficient may be wrong, but the relative sign is as it should
be.) This implies:
\begin{align}
   E_3^{1,1}[3] = &\;({\cal S}\otimes V)/{\cal S}'
   \mbox{ \small\tt generated by } \sigma^{\alpha}
\left((\theta_L\Gamma^m\lambda_L) - (\theta_R\Gamma^m\lambda_R)\right)
\\ 
E_3^{3,0}[3] = &\; S^3{\cal S}/({\cal S}\otimes V)
\end{align}
In the language of $\tilde{E}$, these two components correspond to:
\begin{align}
   (\lambda_L^{\alpha} - \lambda_R^{\alpha})
   \left((\theta_L+\theta_R)\Gamma^m(\lambda_L - \lambda_R)\right)
\label{ThetaLambdaLambda}
\\ 
\mbox{ \small\tt and something cubic in } (\lambda_L - \lambda_R)
\label{SomethingCubic}
\end{align}
There are some equivalence relations. If we contract in (\ref{ThetaLambdaLambda}) the indices
$m$ and $\alpha$ with the gamma-matrix, the resulting expression will be $Q$-exact:
\begin{align}
 &\;     \Gamma_m(\lambda_L + \lambda_R)
   \left((\theta_L+\theta_R)\Gamma^m(\lambda_L + \lambda_R)\right)
\nonumber \\ 
=&\; Q\left(      \Gamma_m(\theta_L + \theta_R)
   \left((\theta_L+\theta_R)\Gamma^m(\lambda_L + \lambda_R)\right)
\right)
\end{align}
There must be a similar equivalence relation in (\ref{SomethingCubic}), which we did not study.

\subsection{Case $N=4$}
\subsubsection{$E_1^{p,q}[4]$ and $d_1$}
\begin{align}
   E_1^{0,0}[4]=\;&0
\\ 
E_1^{1,0}[4]=E_1^{0,1}[4] = E_1^{2,0}[4]  = \;& 0
\\ 
E_1^{1,1}[4] =\;&
({\cal S}\otimes {\cal S}')\oplus ({\cal S}\otimes {\cal S}')
\mbox{ \tt\small generated by }
\nonumber \\ 
& \sigma^{\alpha}(\theta_L^2\lambda_L)_{\beta} \mbox{ \tt\small and }
 \sigma^{\alpha}(\theta_R^2\lambda_R)_{\beta}
\\ 
E_1^{0,2}[4] =\;&
V\otimes V \mbox{ \tt\small generated by }
(\theta_L\Gamma^m\lambda_L) (\theta_R\Gamma^n\lambda_R)
\\ 
E_1^{3,0}[4] = \;& 0
\\ 
E_1^{2,1}[4] = \;& (S^2{\cal S})\otimes (V\oplus V)
\mbox{ \tt\small generated by }
\nonumber \\ 
& \sigma^{\alpha}\sigma^{\beta} (\theta_L\Gamma^m\lambda_L)
\mbox{ \tt\small and }
\sigma^{\alpha}\sigma^{\beta} (\theta_R\Gamma^m\lambda_R)
\\ 
E_1^{1,2}[4] = E_1^{0,3}[4]= \;& 0
\\ 
E_1^{4,0}[4] = \;& S^4 {\cal S} \mbox{ \tt\small generated by }
\sigma^{\alpha}\sigma^{\beta}\sigma^{\gamma}\sigma^{\delta}
\\ 
E_1^{3,1}[4] = E_1^{2,2}[4] = \;& 0
\\ 
E_1^{1,3}[4] = E_1^{0,4}[4] = \;& 0
\end{align}
There is a nontrivial $d_1$:
\begin{align}
   d_1\;:\; E_1^{1,1}[4] \rightarrow & E_1^{2,1}[4]
,\mbox{ \tt\small all other components zero}
\label{D1OnE1114}
\\ 
d_1\left(\sigma^{\alpha}(\theta_L^2\lambda_L)_{\beta}\right) = \;&
\sigma^{\alpha}\left(\sigma\Gamma^m(\theta_L\Gamma_m\lambda_L)\right)_{\beta}
\label{D1OnSigmaThetaL2LambdaL} \\ 
d_1\left(\sigma^{\alpha}(\theta_R^2\lambda_R)_{\beta}\right) = \;&
\sigma^{\alpha}\left(\sigma\Gamma^m(\theta_R\Gamma_m\lambda_R)\right)_{\beta}
\label{D1OnSigmaThetaR2LambdaR}
\end{align}
Notice that $\mbox{ker}\left(d_1:E_1^{1,1}[4] \rightarrow E_1^{2,1}[4]\right) = 0$, therefore $E_2^{1,1}[4] = 0$.

\subsubsection{$E_2^{p,q}$ and $d_2$}
Eq. (\ref{D1OnE1114}) implies that all the components of $E_2[4]$ are the same as the
corresponding components of $E_1[4]$, except for:
\begin{align}
E^{1,1}_2[4] = \;& 0
\\ 
E^{2,1}_2[4] = \;&
{ (S^2{\cal S})\otimes (V\oplus V) \over
  ({\cal S} \otimes {\cal S}')\oplus ({\cal S}\otimes{\cal S}'\;) }
\end{align}
The following components of $d_2\;:\; E_2^{p,q} \rightarrow E_2^{p+2,q-1}$ are potentially nonzero:
\begin{align}
   d_2\;:\;& E_2^{0,2}[4] \rightarrow E_2^{2,1}[4]
\\ 
d_2\;:\;&    E_2^{2,1}[4] \rightarrow E_2^{4,0}[4]
\end{align}

\subsubsection{Calculation of $d_2\;:\;E_2^{0,2}[4] \rightarrow E_2^{2,1}[4]$}\label{sec:CalcD2}
\begin{align}
   d\Big( (\lambda_L\Gamma^m\theta_L) (\lambda_R\Gamma^n\theta_R) \Big)
=\;&
   (\lambda_L\Gamma^m\sigma)(\lambda_R\Gamma^n\theta_R)
+ (\lambda_L\Gamma^m\theta_L)(\lambda_R\Gamma^n\sigma) =
\nonumber \\ 
=\;&
Q\Big(    (\theta_L\Gamma^m\sigma)(\lambda_R\Gamma^n\theta_R)
+ (\theta_R\Gamma^n\sigma)(\lambda_L\Gamma^m\theta_L) \Big)
\end{align}
\begin{align}
 &  d\Big((\theta_L\Gamma^m\sigma)(\lambda_R\Gamma^n\theta_R)
+ (\theta_R\Gamma^n\sigma)(\lambda_L\Gamma^m\theta_L) \Big) =
\nonumber \\ 
=\;& (\sigma\Gamma^m\sigma) (\lambda_R\Gamma^n\theta_R)
- (\sigma\Gamma^n\sigma) (\lambda_L\Gamma^m\theta_L)
- Q\Big( (\sigma\Gamma^m\theta_L)(\theta_R\Gamma^n\sigma) \Big)
\label{D2OnE2024}
\end{align}
In order for the $d_2$ to vanish, this should be in the image of $d_1$, where $d_1$
is given by Eqs. (\ref{D1OnSigmaThetaL2LambdaL}) and (\ref{D1OnSigmaThetaR2LambdaR}). It is immediately clear that $d_2$ annihilates
the scalar component of $E_2^{0,2}[4] = V\otimes V$:
\begin{equation}
   d_2\Big( {\bf C}\subset V\otimes V \Big) = 0
\end{equation}
This means that the cohomology class of $(\theta_L\Gamma^m\lambda_L) (\theta_R\Gamma_m\lambda_R) $ can be
represented by an expression depending only on $\theta_L+\theta_R$, as was demonstrated
in Appendix A2 of {\tt arXiv:1105.2231}.

Also $d_2$ annihilates the antisymmetric tensor:
\begin{equation}
   d_2\Big(
(\lambda_L\Gamma^{[m}\theta_L)(\lambda_R\Gamma^{n]}\theta_R)
\Big) = 0
\end{equation}
which corresponds to (\ref{PhiLLTT2Form}). Indeed:
\begin{equation}
   d_1\Big(
   (\theta_L^2\lambda_L)_{\beta}(\Gamma^{mn})^{\beta}_{\alpha}\sigma^{\alpha}
   \Big) = (\theta_L\Gamma_p\lambda_L) (\sigma\Gamma^p\Gamma^{mn}\sigma)
\simeq (\theta_L\Gamma_{[m}\lambda_L)(\sigma\Gamma_{n]}\sigma)
\end{equation}
and this covers (\ref{D2OnE2024}).

\subsubsection{Calculation of $d_2\;:\;E_2^{2,1}[4] \rightarrow E_2^{4,0}[4]$}
This is essentially the same calculation as $d_2\;:\; E_2^{0,1}[2]\rightarrow E_2^{2,0}[2]$.
We have $E_2^{4,0} = S^4{\cal S}$ and
\begin{equation}
   d_2\left(E_2^{2,1}[4]\right) = S^2{\cal S}\otimes V \subset S^4{\cal S}
\end{equation}
This implies:
\begin{align}
   E_3^{0,2}[4] =\;& {\bf C} \mbox{ \tt\small   (this is  $(\theta_L\Gamma^m\lambda_L) (\theta_R\Gamma_m\lambda_R) $)}
\\ 
E_3^{2,1}[4] =\;&
{(S^2{\cal S})\otimes V\over
  ({\cal S}\otimes ({\cal S}'\oplus {\cal S}'))\;\oplus\;
  (V\otimes V)_0}
\\ 
E_3^{4,0}[4] =\;& {S^4{\cal S}\over S^2{\cal S}\otimes V}
\end{align}
The $d_3$ is zero, therefore $H(d+Q)[4] = E_3[4]$.

Notice that $E_3^{2,1}[4]$ corresponds to something like
$(\lambda_L - \lambda_R)^2 ((\theta_L+\theta_R)\Gamma^m(\lambda_L - \lambda_R))$, and $E_3^{4,0}[4]$ to something quartic in
$\lambda_L - \lambda_R$. In both cases, there are some equivalence relations which we did
not calculate.

\subsubsection{Symbolic computations}
See Section \ref{sec:Computer}. Computation in {\tt L2T2Eqs} show the nontrivial cohomology
class:
\begin{align}
\;&   \lambda_{L+}^{(\bullet}\cap\theta_+\cup\theta_+\cap\lambda_{R+}^{\bullet)}
+  \lambda_{L+}^{(\bullet}\cap\lambda_{R+}\cup\theta_+\cap\theta_+^{\bullet)}
+  \theta_+^{(\bullet}\cap\theta_+\cup\lambda_{L+}\cap\lambda_{R+}^{\bullet)} \;+
\nonumber \\ 
+\;&
\lambda_{R+}^{(\bullet}\cap\theta_+\cup\lambda_{L+}\cap\theta_+^{\bullet)}
- \theta_+^{(\bullet}\cap\lambda_{R+}\cup\lambda_{L+}\cap\theta_+^{\bullet)}
+ \theta_+^{(\bullet}\cap\lambda_{R+}\cup\theta_+\cap\lambda_{L+}^{\bullet)}
\end{align}
This is the part of (\ref{PhiLLTT2Form}) with both indices inside AdS.
Also notice:
\begin{equation}
   \left(\{\lambda_L\capzero\theta\}\cup\{\lambda_R\capzero\theta\} -
\{\lambda_R\capzero\theta\}\cup\{\lambda_L\capzero\theta\}
\right)_{\omega-\mbox{\tt\tiny less}} = 0
\end{equation}

\subsection{Case $N=5$}\label{sec:N5}
\subsubsection{$E_1^{p,q}[5]$}
\begin{align}
   E_1^{0,0}[5] = \;& 0
\\ 
E_1^{1,0}[5] = E_1^{0,1}[5] =\;& 0
\\ 
E_1^{2,0}[5] = E_1^{1,1}[5] =\;& 0
\\ 
E_1^{0,2}[5] =\;& ({\cal S}'\otimes V) \oplus (V\otimes {\cal S}')
\mbox{ \tt\small generated by }
\nonumber \\ 
&
(\theta_L^2\lambda_L)_{\alpha}(\theta_R\lambda_R)^m \;,\;
(\theta_L\lambda_L)^m(\theta_R^2\lambda_R)_{\alpha} \mbox{ \small\tt and }
\nonumber \\ 
&
(\theta_L^3\lambda_L^2)_{\alpha}\;,\; (\theta_R^3\lambda_R^2)_{\alpha}
\\ 
E_1^{3,0}[5] = E_1^{0,3}[5] =\;& 0
\\ 
E_1^{2,1}[5] =\;& S^2{\cal S} \otimes ({\cal S}' \oplus {\cal S}')
\mbox{ \tt\small generated by }
\nonumber\\ 
& \sigma^{\alpha}\sigma^{\beta} (\theta_L^2\lambda_L)_{\gamma}
\mbox{ \small\tt and }
\sigma^{\alpha}\sigma^{\beta} (\theta_R^2\lambda_R)_{\gamma}
\\ 
E_1^{1,2}[5] =\;&
{\cal S}\otimes V\otimes V \mbox{ \small\tt generated by }
\nonumber \\ 
& \sigma^{\alpha}(\theta_L\lambda_L)^m (\theta_R\lambda_R)^n
\\ 
E_1^{4,0}[5] = E_1^{2,2}[5] \;=\;& 0
\\ 
E_1^{1,3}[5] = E_1^{0,4}[5] \;=\;& 0
\\ 
E_1^{3,1}[5] = \;& S^3{\cal S} \otimes (V\oplus V)
\mbox{ \small\tt generated by }
\nonumber\\ 
&
\sigma^{\alpha}\sigma^{\beta}\sigma^{\gamma} (\lambda_L\theta_L)^m
\mbox{ \small\tt and }
\sigma^{\alpha}\sigma^{\beta}\sigma^{\gamma} (\lambda_R\theta_R)^m
\\ 
E_1^{5,0}[5] = \;& S^5{\cal S} \mbox{ \small\tt generated by }
\sigma^{\alpha}\sigma^{\beta}\sigma^{\gamma}\sigma^{\delta}\sigma^{\epsilon}
\\ 
E_1^{4,1}[5] = E_1^{3,2}[5] \; =\;& 0
\\ 
E_1^{2,3}[5] = E_1^{1,4}[5] = E_1^{0,5}[5] \; =\;& 0
\end{align}
\subsubsection{$\lambda^2\theta^3$}
\paragraph     {Calculation of $E_2^{2,1}[5]$.}
The map $d_1:E_2^{2,1}[5] \rightarrow E_2^{3,1}[5]$ is given by:
\begin{align}
   d_1\Big( \sigma^{\alpha}\sigma^{\beta} (\theta_L^2\lambda_L)_{\gamma}\Big)
\simeq \sigma^{\alpha}\sigma^{\beta} (\Gamma^m\sigma)_{\gamma}
(\theta_L\Gamma_m\lambda_L)
\\ 
   d_1\Big( \sigma^{\alpha}\sigma^{\beta} (\theta_R^2\lambda_R)_{\gamma}\Big)
\simeq \sigma^{\alpha}\sigma^{\beta} (\Gamma^m\sigma)_{\gamma}
(\theta_R\Gamma_m\lambda_R)
\end{align}
We observe that $\mbox{ker }(d_1:E_1^{2,1}[5]\rightarrow E_1^{3,1}[5])$ is generated by the
following classes:
\begin{align}\label{E112}
(\lambda_L\Gamma^p\theta_L)(\theta_L\Gamma_p\Gamma_{mn}\sigma)\;\Gamma^{mn}\sigma
+ A(\lambda_L\Gamma^p\theta_L)(\theta_L\Gamma_p\sigma)\;\sigma
\end{align}
and the same class but with $\lambda_L \leftrightarrow \lambda_R$, where $A$ is the coefficient such
that\footnote{Notice that $(\sigma\Gamma_p\Gamma_{klmn}\sigma)\Gamma^{klmn}\sigma$ is
linearly independent from $(\sigma\Gamma_p\sigma)\sigma$, therefore there is
no such term in (\ref{E112})}
\begin{equation}
(\sigma\Gamma_p\Gamma_{mn}\sigma)\;\Gamma^{mn}\sigma
+ A(\sigma\Gamma_p\sigma)\;\sigma = 0
\end{equation}
Since $E_1^{1,1}[5] = 0$, we conclude that $E_2^{2,1}[5] = S\oplus S$ generated by (\ref{E112}) and
$\lambda_L\leftrightarrow \lambda_R$.

\paragraph     {Calculation of $E_2^{0,2}[5]$ and  $E_2^{1,2}[5]$.}
The kernel $\mbox{ker}\;d_1: E_1^{0,2}[5]\rightarrow E_1^{1,2}[5]$ is generated by:
\begin{align}\label{KerD1E102}
& (\theta_L^3\lambda_L^2)_{\alpha} \;,\;\; (\theta_R^3\lambda_R^2)_{\alpha}
\end{align}
This means that $E_2^{0,2}[5]$ is generated by (\ref{KerD1E102}).

Also notice that $\mbox{im } d_1: E_1^{0,2}[5]\rightarrow E_1^{1,2}[5]$ is generated by:
\begin{align}
&(\theta_R\Gamma^m\lambda_R)(\theta_L\Gamma^n\lambda_L)\Gamma_{mn}\sigma
\label{ZeroInE2125a}\\ 
   \;\mbox{ \tt\small and }\;&
(\theta_R\Gamma^m\lambda_R)(\theta_L\Gamma_m\lambda_L)\sigma
\label{ZeroInE2125b}
\end{align}
and that $d_1: E_1^{1,2}[5]\rightarrow E_1^{2,2}[5]$ is zero because $E_1^{2,2}[5]=0$. This means that
$E_2^{1,2}[5]$ is generated by $(\theta_L\Gamma^m\lambda_L)(\theta_R\Gamma^n\lambda_R)\sigma^{\alpha}$ modulo (\ref{ZeroInE2125a}) and (\ref{ZeroInE2125b}).

\paragraph     {Calculation of $d_2: E_2^{0,2}[5]\to E_2^{2,1}[5]$.}
It turns out that both $(\theta_L^3\lambda_L^2)_{\alpha}$
and  $(\theta_R^3\lambda_R^2)_{\alpha}$ are acted upon nontrivially by $d_2$, and therefore do not survive
in $E_3$. Notice that $d_2(\theta_L^3\lambda_L^2)$ is necessarily proportional to (\ref{E112}). This
implies that
\begin{itemize}
\item $E_2^{0,2}[5]$ is cancelled against $E_2^{2,1}[5]$, {\it i.e.} $E_3^{0,2}[5] = E_3^{2,1}[5] = 0$
\end{itemize}
Vanishing of $E_3^{0,2}[5]$ implies that there is no cohomology of the type $\lambda^2\theta^3$.
This can be demonstrated explicitly, in the following way. The leading term
in $\lambda_L$ would be $(\lambda_L\Gamma_m\theta)(\lambda_L\Gamma_n\theta)\Gamma_{mn}\theta$. But this is not annihilated by $Q_R$.
We used a computer calculation ({\tt L2T3Eqs} in Section \ref{sec:Computer}) to confirm that
indeed there is no cohomology of the type $\lambda^2\theta^3$.

\paragraph     {Calculation of $E_3^{1,2}$}
This is parallel to Section \ref{sec:CalcD2}:
\begin{align}
   d\Big( (\lambda_L\Gamma^m\theta_L) (\lambda_R\Gamma^n\theta_R) \sigma^{\alpha}\Big)
=\;&
   (\lambda_L\Gamma^m\sigma)(\lambda_R\Gamma^n\theta_R) \sigma^{\alpha}
+ (\lambda_L\Gamma^m\theta_L)(\lambda_R\Gamma^n\sigma) \sigma^{\alpha} =
\nonumber \\ 
=\;&
Q\Big(    (\theta_L\Gamma^m\sigma)(\lambda_R\Gamma^n\theta_R)\sigma^{\alpha}
+ (\theta_R\Gamma^n\sigma)(\lambda_L\Gamma^m\theta_L)\sigma^{\alpha}  \Big)
\end{align}
\begin{align}
 &  d\Big((\theta_L\Gamma^m\sigma)(\lambda_R\Gamma^n\theta_R)\sigma^{\alpha}
+ (\theta_R\Gamma^n\sigma)(\lambda_L\Gamma^m\theta_L)\sigma^{\alpha}  \Big) =
\nonumber \\ 
=\;& \sigma^{\alpha}(\sigma\Gamma^m\sigma) (\lambda_R\Gamma^n\theta_R)
- \sigma^{\alpha}(\sigma\Gamma^n\sigma) (\lambda_L\Gamma^m\theta_L)
- Q\Big( \sigma^{\alpha}(\sigma\Gamma^m\theta_L)(\theta_R\Gamma^n\sigma) \Big)
\end{align}
By the same argument as in Section \ref{sec:CalcD2}, the computation of the kernel of
$d_2: E_2^{1,2}[5] \rightarrow E_2^{3,1}[5]$ is equivalent to the problem of intertwining:
\begin{align}
\;&   \sigma^{\alpha}\;\sigma^{\gamma}\; (a_{(L)}^m\Gamma_m\sigma)_{\beta}\;\;,\;\;\;
\sigma^{\alpha}\;\sigma^{\gamma}\; (a_{(R)}^m\Gamma_m\sigma)_{\beta}
\label{AtoIntertwine}\\ 
\mbox{\tt\small with }\;& \sigma^{\alpha} \left(
   (\sigma\Gamma^m\sigma) a^{(R)n} - (\sigma\Gamma^n\sigma) a^{(R)m}
\right)
\label{BtoIntertwine}
\end{align}
(where $a_{(L)}^n$ stands for $\lambda_L\Gamma^n\theta_L$ and $a_{(R)}^n$ for $\lambda_L\Gamma^n\theta_L$). This is only possible
with one of the following options:
\begin{enumerate}
   \item $m$ and $n$ in (\ref{BtoIntertwine}) are contracted
   \item $m$ and $n$ in (\ref{BtoIntertwine}) are antisymmetrized
   \item $\alpha$ and $n$ in (\ref{BtoIntertwine}) are contracted through a $\Gamma$-matrix
\end{enumerate}
But the first and the third options are not interesting, because the
corresponding element of $E_2^{1,2}[5]$ is actually zero, {\it i.e.} in the image
of $d_1: E_1^{0,2}[5]\rightarrow E_1^{1,2}[5]$. Therefore we conclude that
\begin{itemize}
\item $E_3^{1,2}[5]$ is generated by:
\begin{align}
(\lambda_L\Gamma^{[m}\theta_L)(\lambda_R\Gamma^{n]}\theta_R)\sigma^{\alpha}
\end{align}
This corresponds to (\ref{Lambda3Theta2}).
\end{itemize}
The following expression is an example of a nontrivial class:
\begin{align}
&   \lambda_{R+}\cup\theta_+\cap\lambda_{L+}\cup\theta_+\cap\lambda_{R+}
+ {1\over 4}||\lambda_{R+}\cup\lambda_{R+}|| \theta_+\cup\lambda_{L+}\cap\theta_+ -
\nonumber \\ 
& - {1\over 4}||\lambda_{R+}\cup\lambda_{R+}|| (\theta_+\cup\theta_+\cap\lambda_{R+} +
\lambda_{R+}\cup\theta_+\cap\theta_+)
\label{MTLTM}
\end{align}
We verified by a computer calculation ({\tt L3T2Eqs} in Section \ref{sec:Computer}) that this
is not BRST-exact.

\subsection{Case $N=6$}
\begin{align}
   E_1^{0,0}[6] =\;& 0
\\ 
E_1^{1,0}[6] = E_1^{0,1}[6] =\;& 0
\\ 
E_1^{2,0}[6] = E_1^{1,1}[6] =\;& 0
\\ 
E_1^{0.2}[6] =\;& (\theta_L^2\lambda_L)_{\alpha} (\theta_R^2\lambda_R)_{\beta}
\;,\; \nonumber\\ 
&
(\theta_L^4\lambda^2_L)^m\mbox{ \tt\small and } (\theta_R^4\lambda^2_R)^m
\\ 
E_1^{3,0}[6] = E_1^{2,1}[6] = E_1^{0,3}[6] =\;& 0
\\ 
E_1^{1,2}[6] = \;&
\sigma^{\alpha} (\theta_L\lambda_L)^m (\theta_R^2\lambda_R)_{\beta}
\nonumber\\ 
&
\sigma^{\alpha} (\theta^2_L\lambda_L)_{\beta} (\theta_R\lambda_R)^m
\nonumber \\ 
&
\sigma^{\alpha} (\theta^3_L\lambda^2_L)^{\beta}
\mbox{ \tt\small and }
\sigma^{\alpha} (\theta^3_R\lambda^2_R)^{\beta}
\\ 
E_1^{4,0}[6] = E_1^{1,3}[6] = E_1^{0,4}[6] = \;& 0
\\ 
E_1^{3,1}[6] = \;&
\sigma^{\alpha}\sigma^{\beta}\sigma^{\gamma}(\theta_L^2\lambda_L)_{\delta}
\mbox{ \tt\small and }
\sigma^{\alpha}\sigma^{\beta}\sigma^{\gamma}(\theta_R^2\lambda_R)_{\delta}
\\ 
E_1^{2,2}[6] = \;&
\sigma^{\alpha}\sigma^{\beta}(\theta_L\lambda_L)^m (\theta_R\lambda_R)^n
\\ 
E_1^{5,0}[6] = E_1^{3,2}[6]= E_1^{2,3}[6] = \;& 0
\\ 
E_1^{1,4}[6] = E_1^{0,5}[6] = \;& 0
\\ 
E_1^{4,1}[6] = \;& \sigma^{\alpha}\sigma^{\beta}\sigma^{\gamma}\sigma^{\delta}
(\theta_L\lambda_L)^m \mbox{ \tt\small and }
\nonumber \\ 
&
\sigma^{\alpha}\sigma^{\beta}\sigma^{\gamma}\sigma^{\delta}
(\theta_R\lambda_R)^m
\\ 
E_1^{6,0}[6] = \;&  \sigma^{\alpha_1}\cdots\sigma^{\alpha_6}
\\ 
E_1^{5,1}[6] = E_1^{4,2} = E_1^{3,3} = \; & 0
\\ 
E_1^{2,4}[6] = E_1^{1,5} = E_1^{0,6} = \; & 0
\end{align}
In fact $E_1^{0,2}[6]$ does not survive higher order corrections. Indeed, suppose
that there is a nontrivial cohomology class of the form $\lambda^2\theta^4$. It should have
the form:
\begin{equation}
   \lambda_L^2\theta^4 + \ldots
\end{equation}
where $\ldots$ stands for terms containing $\lambda_L\lambda_R\theta^4$ and $\lambda_R^2\theta^4$. The derivative
of such an expression with respect to $\theta$ would have started with $\lambda_L^2\theta^3$
representing a nontrivial cohomology class of the type $\lambda^2\theta^3$. But there is no
such class --- see Section \ref{sec:N5}. We have confirmed by a symbolic computation
that there is no such class ({\tt L2T4Eqs} in Section \ref{sec:Computer}).

Now consider a class of the order $\lambda^3\theta^3$. Then the leading term in $\lambda_L$ should be
a linear combination of the following two:
\begin{align}
\;&
\theta^{\alpha}_+\cap\lambda_{L+}
\cup\theta_+\cap\theta_+\cup\lambda_{L+}\cap\lambda^{\beta}_{R+}
\\ 
\mbox{\tt\small and }\;&
\lambda^{\alpha}_{R+}\cap\lambda_{L+}\cup\theta_+\cap\theta_+
\cup\lambda_{L+}\cap\theta^{\beta}_+
\end{align}
Both can be made  $Q^{[0]}$-closed by adding the terms of the type $\lambda_{R+}^2\lambda_{L+}\theta^3$ and
$\lambda_{R+}^3\theta^3$.  The part antisymmetric in $\alpha\leftrightarrow\beta$ then becomes $Q^{[0]}$-exact:
\begin{align}
\;& - \theta^{\alpha}_+\cap\lambda_{L+}
\cup\theta_+\cap\theta_+\cup\lambda_{L+}\cap\lambda^{\beta}_{R+} \;+\;
\lambda^{\alpha}_{R+}\cap\lambda_{L+}\cup\theta_+\cap\theta_+
\cup\lambda_{L+}\cap\theta^{\beta}_+ \;=
\nonumber \\ 
=\;&
\left(
   \lambda_{L+}{\partial\over\partial\theta_+} +
   \lambda_{R+}{\partial\over\partial\theta_+}
\right)\Big(
\;{5\over 8}\;\theta^{\alpha}_+\cap\lambda_{L+}
\cup\theta_+\cap\theta_+\cup\lambda_{L+}\cap\theta^{\beta}_{+} +
\nonumber \\
&\qquad \qquad\qquad \qquad \qquad
+\; \mbox{\tt\small some }[\lambda_L\lambda_R\theta^4] +
\;\mbox{\tt\small some }[\lambda_R^2\theta^4]
\Big)
\end{align}
But the part symmetric in $(\alpha\leftrightarrow\beta)$ does represent a nontrivial cohomology
class. There must be also a class of the type (\ref{PhiL3T3Mixed1}) and (\ref{PhiL3T3Mixed2}),
but we have not verified it.

The general theory behind the classed of the order $\lambda^3\theta^3$ is the following. We
explained in Section \ref{sec:N5} that $d_1$ acts nontrivially on $(\theta_L^2\lambda_L)(\theta_R\lambda_R)$ and
$(\theta_R^2\lambda_R)(\theta_L\lambda_L)$. This was the reason why they do not survive in $E_2[5]$. However,
let us now consider the following ${\bf Z}_2^{LR}$-odd element of $E^{1,2}_1[6]$:
\begin{align}
(\sigma \Gamma_{abc} \Gamma_n\Gamma_m\theta_L)
(\theta_L\Gamma^m\lambda_L)(\theta_R\Gamma^n\lambda_R)
+ (L\leftrightarrow R)
\label{SigmaThetaL2ThetaRLambdaLLambdaR}
\end{align}
The $d_1$ annihilates this element. Indeed, the $d_1$ of this element is necessarily
proportional to:
\begin{equation}
   (\sigma\Gamma_{abc}\Gamma_{mn}\sigma)
(\theta_L\Gamma^m\lambda_L)(\theta_R\Gamma^n\lambda_R)
\end{equation}
which is ${\bf Z}_2^{LR}$-even; therefore $d_1$ of (\ref{SigmaThetaL2ThetaRLambdaLLambdaR}) is zero.

The $d_1$ of (\ref{SigmaThetaL2ThetaRLambdaLLambdaR}) being zero implies that the $d_2$ of it is well defined. The
value of $d_2$ on (\ref{SigmaThetaL2ThetaRLambdaLLambdaR}) is of the form:
\begin{equation}\label{FormOfD2OnE121}
   [\sigma^3\lambda_L\theta_L^2] + [\sigma^3\lambda_R\theta_R^2]
\end{equation}
On the other hand, let us consider another ${\bf Z}_2^{LR}$-odd element, of the form:
\begin{align}
(\sigma \Gamma_{abc} \Gamma_n\Gamma_m\theta_L)(\theta_L\Gamma^m\lambda_L)
(\theta_L\Gamma^n\lambda_L) + (L\leftrightarrow R)
\label{SigmaThetaL3LambdaL2}
\end{align}
The $d_2$ of (\ref{SigmaThetaL3LambdaL2}) is of the same form as the $d_2$ of (\ref{SigmaThetaL3LambdaL2}),
namely of the form (\ref{FormOfD2OnE121}). Therefore:
\begin{itemize}
\item some linear combination of (\ref{SigmaThetaL2ThetaRLambdaLLambdaR}) and (\ref{SigmaThetaL3LambdaL2}) survives on $E^{1,2}_{\infty}[6]$
\end{itemize}
This gives (\ref{PhiL3T3}).

\subsection{$N=7$}
There are no nontrivial classes of the form $\lambda^2\theta^5$. To prove this, we have to
consider the term with the highest number of $\lambda_L$. It has to be annihilated by
$\lambda_L{\partial\over\partial\theta}$. Because the cohomology of $\lambda_L{\partial\over\partial\theta}$ in degrees $\lambda_L^2\theta^5$, $\lambda_L\theta^5$ and $\lambda_L^0\theta^5$
is zero, this can be always gauged away.

However, there is a nontrivial class of the form $\lambda^3\theta^4$ --- see (\ref{TypeL3T4}) or (\ref{TypeL3T4Gamma}).

\subsection{$N=8$}
At the level eight we have a nontrivial $E_1^{0,3}[8]$ generated by $\lambda_L^3\theta_L^5$, $\lambda_R^3\theta_R^5$,
$(\lambda_L^2\theta_L^4)(\lambda_R\theta_R)$, $(\lambda_R^2\theta_R^4)(\lambda_L\theta_L)$, $(\lambda_L^2\theta_L^3)(\lambda_R\theta_R^2)$, $(\lambda_R^2\theta_R^3)(\lambda_L\theta_L^2)$. The classes
surviving on $E_2$ ({\it i.e.} annihilated by the $d_1$) are:
\begin{itemize}
\item $\lambda_L^3\theta_L^5$ and $\lambda_R^3\theta_R^5$
\item some linear combinations of the type:
\begin{equation}
a(\lambda_L^2\theta_L^4)(\lambda_R\theta_R)
+ b(\lambda_L^2\theta_L^3)(\lambda_R\theta_R^2)
\;\mbox{ \small\tt and }\;
a(\lambda_R^2\theta_R^4)(\lambda_L\theta_L) + b(\lambda_R^2\theta_R^3)(\lambda_L\theta_L^2)
\label{N8SomeLinearCombinations}
\end{equation}
\end{itemize}
What happens when we pass to $E_3$? Consider $d_2:E_2^{0,3}[8] \rightarrow E_2^{2,2}[8]$. The
potential obstacle is in $E_2^{2,2}[8]$.

Let us look at $E_1^{2,2}[8]$:
\begin{align}
   E_1^{2,2}[8]\;:\; &
   \sigma^{\alpha}\sigma^{\beta}(\lambda_L^2\theta_L^4)^m
   \mbox{ \tt\small and }
   \sigma^{\alpha}\sigma^{\beta}(\lambda_R^2\theta_R^4)^m
   \mbox{ \tt\small and }
   \sigma^{\alpha}\sigma^{\beta}(\lambda_L\theta_L^2)_{\gamma}
   (\lambda_R\theta_R^2)_{\delta}
\end{align}
Also notice that $E_1^{1,2}[8]=0$, and therefore nothing in $E_1^{2,2}[8]$ is in the image
of $d_1$. We will now use the  symmetry ${\bf Z}_2^{LR}$ which was described in Section \ref{sec:Bicomplex}.
Let us look at those elements of $E^{2,2}_1[8]$ which are scalars and are of the
type $[\sigma^2\lambda_L\lambda_R\theta_L^2\theta_R^2]$. They are coming
from $\sigma^{\alpha}\sigma^{\beta}(\lambda_L\theta_L^2)_{\gamma}
(\lambda_R\theta_R^2)_{\delta}$. They are all even\footnote{Indeed, the odd elements
would be those containing $\lambda_L\theta_L^2$ and $\lambda_R\theta_R^2$ through $((\lambda_L\theta_L^2)\Gamma_{pqr}(\lambda_R\theta_R^2))$. But
$\sigma^{\alpha}\sigma^{\beta}$ is symmetric in $\alpha\leftrightarrow\beta$ and therefore does not contain a three-form.} under  ${\bf Z}_2^{LR}$. We therefore avoid
this obstacle,  if we simply restrict ourselves to  ${\bf Z}_2^{LR}$-odd elements.

It is enough to get rid of the obstacles $\sigma^{\alpha}\sigma^{\beta}(\lambda_L^2\theta_L^4)^m$ and $\sigma^{\alpha}\sigma^{\beta}(\lambda_R^2\theta_R^4)^m$.
Let us consider the ${\bf Z}_2^{LR}$-odd combination of the form:
\begin{align}
  & \Big( (\lambda_L^3\theta_L^5)
+ a (\lambda_L^2\theta_L^4)^m(\lambda_R\theta_R)^m
+ b (\lambda_L^2\theta_L^3)^{\alpha}(\lambda_R\theta_R^2)_{\alpha} \Big) \; -
\nonumber
\\ 
-\;& \Big(
(\lambda_R^3\theta_R^5) +
a (\lambda_R^2\theta_R^4)^m(\lambda_L\theta_L)^m +
b (\lambda_R^2\theta_R^3)^{\alpha}(\lambda_L\theta_L^2)_{\alpha}
\Big)
\label{L3T5LR}
\end{align}
The ratio of the coefficients $a$ and $b$ is fixed as in (\ref{N8SomeLinearCombinations}), in other words:
\begin{align}
  & d_1\Big((\lambda_L^3\theta_L^5)
+ a (\lambda_L^2\theta_L^4)^m(\lambda_R\theta_R)^m
+ b (\lambda_L^2\theta_L^3)^{\alpha}(\lambda_R\theta_R^2)_{\alpha} \Big) \; = \;0
\end{align}
Now let us adjust $a$ and $b$ (keeping their ratio) so that:
\begin{align}
& d_2\Big((\lambda_L^3\theta_L^5)
+ a (\lambda_L^2\theta_L^4)^m(\lambda_R\theta_R)^m
+ b (\lambda_L^2\theta_L^3)^{\alpha}(\lambda_R\theta_R^2)_{\alpha} \Big) \; = \;
\\ 
=\; & c_1 (\sigma \Gamma^m \sigma)
\left((\lambda_L\theta_L^2) \Gamma^m (\lambda_R\theta_R^2)\right)
+ c_5 (\sigma \Gamma^{m_1\ldots m_5} \sigma)
\left((\lambda_L\theta_L^2) \Gamma^{m_1\ldots m_5} (\lambda_R\theta_R^2)\right)
\nonumber
\end{align}
with some $c_1$ and $c_2$; the point is that it is possible to choose $a$ and $b$ so
that the $d_2(\ldots)$ does not have any terms proportional to $\sigma\sigma(\lambda_L^2\theta_L^4)$. (Notice
that the absence of the terms proportional to $\sigma\sigma(\lambda_R^2\theta_R^4)$ is automatic due to
the ${\bf Z}_2^{LR}$-symmetry.) This means that we removed the obstacles
$\sigma^{\alpha}\sigma^{\beta}(\lambda_L^2\theta_L^4)^m$ and $\sigma^{\alpha}\sigma^{\beta}(\lambda_R^2\theta_R^4)^m$.
\begin{align}
   d_2\;&\;\Big[\phantom{-}
\Big( (\lambda_L^3\theta_L^5)
+ a (\lambda_L^2\theta_L^4)^m(\lambda_R\theta_R)^m
+ b (\lambda_L^2\theta_L^3)^{\alpha}(\lambda_R\theta_R^2)_{\alpha} \Big) \; -
\nonumber
\\ 
& \;-\Big(
(\lambda_R^3\theta_R^5) +
a (\lambda_R^2\theta_R^4)^m(\lambda_L\theta_L)^m +
b (\lambda_R^2\theta_R^3)^{\alpha}(\lambda_L\theta_L^2)_{\alpha}
\Big)\;\;\Big] = 0
\end{align}
Furthermore:
\begin{align}
   d_3\;:\; E_3^{0,3}[8]\to E_3^{3,1}[8] = &0
\\ 
   d_4\;:\; E_4^{0,3}[8]\to E_4^{4,0}[8] = &0
\end{align}
We conclude that with this special choice of $a$ and $b$ (\ref{L3T5LR}) actually survives
all the way up to $E_{\infty}$. This means that there must be a representative
function of $\theta_L + \theta_R$; this is Eq. (\ref{TypeL3T5Gamma}).

\section*{Acknowledgments}
We would like to thank N.J.~Berkovits and A.S.~Schwarz for many useful
suggestions. This work was supported in part by the Ministry of Education and
Science of the Russian Federation under the project 14.740.11.0347
``Integrable and algebro-geometric structures in string theory and quantum
field theory'', and in part by the RFFI grant 10-02-01315
``String theory and integrable systems''.

% \bibliographystyle{JHEP} \renewcommand{\refname}{Bibliography}
% \addcontentsline{toc}{section}{Bibliography}
% \bibliography{../andrei}

\def\cprime{$'$} \def\cprime{$'$}
\providecommand{\href}[2]{#2}\begingroup\raggedright\endgroup

\end{document}